\documentclass[usenatbib,usegraphicx]{mn2e}
\usepackage{comment}
\usepackage{amsfonts}
\usepackage{amsmath}
\usepackage{amssymb}
\usepackage{color}
\definecolor{AliceBlue}{rgb}{0.94,0.97,1.00}
\definecolor{AntiqueWhite1}{rgb}{1.00,0.94,0.86}
\definecolor{AntiqueWhite2}{rgb}{0.93,0.87,0.80}
\definecolor{AntiqueWhite3}{rgb}{0.80,0.75,0.69}
\definecolor{AntiqueWhite4}{rgb}{0.55,0.51,0.47}
\definecolor{AntiqueWhite}{rgb}{0.98,0.92,0.84}
\definecolor{BlanchedAlmond}{rgb}{1.00,0.92,0.80}
\definecolor{BlueViolet}{rgb}{0.54,0.17,0.89}
\definecolor{CadetBlue1}{rgb}{0.60,0.96,1.00}
\definecolor{CadetBlue2}{rgb}{0.56,0.90,0.93}
\definecolor{CadetBlue3}{rgb}{0.48,0.77,0.80}
\definecolor{CadetBlue4}{rgb}{0.33,0.53,0.55}
\definecolor{CadetBlue}{rgb}{0.37,0.62,0.63}
\definecolor{CornflowerBlue}{rgb}{0.39,0.58,0.93}
\definecolor{DarkBlue}{rgb}{0.00,0.00,0.55}
\definecolor{DarkCyan}{rgb}{0.00,0.55,0.55}
\definecolor{DarkGoldenrod1}{rgb}{1.00,0.73,0.06}
\definecolor{DarkGoldenrod2}{rgb}{0.93,0.68,0.05}
\definecolor{DarkGoldenrod3}{rgb}{0.80,0.58,0.05}
\definecolor{DarkGoldenrod4}{rgb}{0.55,0.40,0.03}
\definecolor{DarkGoldenrod}{rgb}{0.72,0.53,0.04}
\definecolor{DarkGray}{rgb}{0.66,0.66,0.66}
\definecolor{DarkGreen}{rgb}{0.00,0.39,0.00}
\definecolor{DarkGrey}{rgb}{0.66,0.66,0.66}
\definecolor{DarkKhaki}{rgb}{0.74,0.72,0.42}
\definecolor{DarkMagenta}{rgb}{0.55,0.00,0.55}
\definecolor{DarkOliveGreen1}{rgb}{0.79,1.00,0.44}
\definecolor{DarkOliveGreen2}{rgb}{0.74,0.93,0.41}
\definecolor{DarkOliveGreen3}{rgb}{0.64,0.80,0.35}
\definecolor{DarkOliveGreen4}{rgb}{0.43,0.55,0.24}
\definecolor{DarkOliveGreen}{rgb}{0.33,0.42,0.18}
\definecolor{DarkOrange1}{rgb}{1.00,0.50,0.00}
\definecolor{DarkOrange2}{rgb}{0.93,0.46,0.00}
\definecolor{DarkOrange3}{rgb}{0.80,0.40,0.00}
\definecolor{DarkOrange4}{rgb}{0.55,0.27,0.00}
\definecolor{DarkOrange}{rgb}{1.00,0.55,0.00}
\definecolor{DarkOrchid1}{rgb}{0.75,0.24,1.00}
\definecolor{DarkOrchid2}{rgb}{0.70,0.23,0.93}
\definecolor{DarkOrchid3}{rgb}{0.60,0.20,0.80}
\definecolor{DarkOrchid4}{rgb}{0.41,0.13,0.55}
\definecolor{DarkOrchid}{rgb}{0.60,0.20,0.80}
\definecolor{DarkRed}{rgb}{0.55,0.00,0.00}
\definecolor{DarkSalmon}{rgb}{0.91,0.59,0.48}
\definecolor{DarkSeaGreen1}{rgb}{0.76,1.00,0.76}
\definecolor{DarkSeaGreen2}{rgb}{0.71,0.93,0.71}
\definecolor{DarkSeaGreen3}{rgb}{0.61,0.80,0.61}
\definecolor{DarkSeaGreen4}{rgb}{0.41,0.55,0.41}
\definecolor{DarkSeaGreen}{rgb}{0.56,0.74,0.56}
\definecolor{DarkSlateBlue}{rgb}{0.28,0.24,0.55}
\definecolor{DarkSlateGray1}{rgb}{0.59,1.00,1.00}
\definecolor{DarkSlateGray2}{rgb}{0.55,0.93,0.93}
\definecolor{DarkSlateGray3}{rgb}{0.47,0.80,0.80}
\definecolor{DarkSlateGray4}{rgb}{0.32,0.55,0.55}
\definecolor{DarkSlateGray}{rgb}{0.18,0.31,0.31}
\definecolor{DarkSlateGrey}{rgb}{0.18,0.31,0.31}
\definecolor{DarkTurquoise}{rgb}{0.00,0.81,0.82}
\definecolor{DarkViolet}{rgb}{0.58,0.00,0.83}
\definecolor{DeepPink1}{rgb}{1.00,0.08,0.58}
\definecolor{DeepPink2}{rgb}{0.93,0.07,0.54}
\definecolor{DeepPink3}{rgb}{0.80,0.06,0.46}
\definecolor{DeepPink4}{rgb}{0.55,0.04,0.31}
\definecolor{DeepPink}{rgb}{1.00,0.08,0.58}
\definecolor{DeepSkyBlue1}{rgb}{0.00,0.75,1.00}
\definecolor{DeepSkyBlue2}{rgb}{0.00,0.70,0.93}
\definecolor{DeepSkyBlue3}{rgb}{0.00,0.60,0.80}
\definecolor{DeepSkyBlue4}{rgb}{0.00,0.41,0.55}
\definecolor{DeepSkyBlue}{rgb}{0.00,0.75,1.00}
\definecolor{DimGray}{rgb}{0.41,0.41,0.41}
\definecolor{DimGrey}{rgb}{0.41,0.41,0.41}
\definecolor{DodgerBlue1}{rgb}{0.12,0.56,1.00}
\definecolor{DodgerBlue2}{rgb}{0.11,0.53,0.93}
\definecolor{DodgerBlue3}{rgb}{0.09,0.45,0.80}
\definecolor{DodgerBlue4}{rgb}{0.06,0.31,0.55}
\definecolor{DodgerBlue}{rgb}{0.12,0.56,1.00}
\definecolor{FloralWhite}{rgb}{1.00,0.98,0.94}
\definecolor{ForestGreen}{rgb}{0.13,0.55,0.13}
\definecolor{GhostWhite}{rgb}{0.97,0.97,1.00}
\definecolor{GreenYellow}{rgb}{0.68,1.00,0.18}
\definecolor{HotPink1}{rgb}{1.00,0.43,0.71}
\definecolor{HotPink2}{rgb}{0.93,0.42,0.65}
\definecolor{HotPink3}{rgb}{0.80,0.38,0.56}
\definecolor{HotPink4}{rgb}{0.55,0.23,0.38}
\definecolor{HotPink}{rgb}{1.00,0.41,0.71}
\definecolor{IndianRed1}{rgb}{1.00,0.42,0.42}
\definecolor{IndianRed2}{rgb}{0.93,0.39,0.39}
\definecolor{IndianRed3}{rgb}{0.80,0.33,0.33}
\definecolor{IndianRed4}{rgb}{0.55,0.23,0.23}
\definecolor{IndianRed}{rgb}{0.80,0.36,0.36}
\definecolor{LavenderBlush1}{rgb}{1.00,0.94,0.96}
\definecolor{LavenderBlush2}{rgb}{0.93,0.88,0.90}
\definecolor{LavenderBlush3}{rgb}{0.80,0.76,0.77}
\definecolor{LavenderBlush4}{rgb}{0.55,0.51,0.53}
\definecolor{LavenderBlush}{rgb}{1.00,0.94,0.96}
\definecolor{LawnGreen}{rgb}{0.49,0.99,0.00}
\definecolor{LemonChiffon1}{rgb}{1.00,0.98,0.80}
\definecolor{LemonChiffon2}{rgb}{0.93,0.91,0.75}
\definecolor{LemonChiffon3}{rgb}{0.80,0.79,0.65}
\definecolor{LemonChiffon4}{rgb}{0.55,0.54,0.44}
\definecolor{LemonChiffon}{rgb}{1.00,0.98,0.80}
\definecolor{LightBlue1}{rgb}{0.75,0.94,1.00}
\definecolor{LightBlue2}{rgb}{0.70,0.87,0.93}
\definecolor{LightBlue3}{rgb}{0.60,0.75,0.80}
\definecolor{LightBlue4}{rgb}{0.41,0.51,0.55}
\definecolor{LightBlue}{rgb}{0.68,0.85,0.90}
\definecolor{LightCoral}{rgb}{0.94,0.50,0.50}
\definecolor{LightCyan1}{rgb}{0.88,1.00,1.00}
\definecolor{LightCyan2}{rgb}{0.82,0.93,0.93}
\definecolor{LightCyan3}{rgb}{0.71,0.80,0.80}
\definecolor{LightCyan4}{rgb}{0.48,0.55,0.55}
\definecolor{LightCyan}{rgb}{0.88,1.00,1.00}
\definecolor{LightGoldenrod1}{rgb}{1.00,0.93,0.55}
\definecolor{LightGoldenrod2}{rgb}{0.93,0.86,0.51}
\definecolor{LightGoldenrod3}{rgb}{0.80,0.75,0.44}
\definecolor{LightGoldenrod4}{rgb}{0.55,0.51,0.30}
\definecolor{LightGoldenrodYellow}{rgb}{0.98,0.98,0.82}
\definecolor{LightGoldenrod}{rgb}{0.93,0.87,0.51}
\definecolor{LightGray}{rgb}{0.83,0.83,0.83}
\definecolor{LightGreen}{rgb}{0.56,0.93,0.56}
\definecolor{LightGrey}{rgb}{0.83,0.83,0.83}
\definecolor{LightPink1}{rgb}{1.00,0.68,0.73}
\definecolor{LightPink2}{rgb}{0.93,0.64,0.68}
\definecolor{LightPink3}{rgb}{0.80,0.55,0.58}
\definecolor{LightPink4}{rgb}{0.55,0.37,0.40}
\definecolor{LightPink}{rgb}{1.00,0.71,0.76}
\definecolor{LightSalmon1}{rgb}{1.00,0.63,0.48}
\definecolor{LightSalmon2}{rgb}{0.93,0.58,0.45}
\definecolor{LightSalmon3}{rgb}{0.80,0.51,0.38}
\definecolor{LightSalmon4}{rgb}{0.55,0.34,0.26}
\definecolor{LightSalmon}{rgb}{1.00,0.63,0.48}
\definecolor{LightSeaGreen}{rgb}{0.13,0.70,0.67}
\definecolor{LightSkyBlue1}{rgb}{0.69,0.89,1.00}
\definecolor{LightSkyBlue2}{rgb}{0.64,0.83,0.93}
\definecolor{LightSkyBlue3}{rgb}{0.55,0.71,0.80}
\definecolor{LightSkyBlue4}{rgb}{0.38,0.48,0.55}
\definecolor{LightSkyBlue}{rgb}{0.53,0.81,0.98}
\definecolor{LightSlateBlue}{rgb}{0.52,0.44,1.00}
\definecolor{LightSlateGray}{rgb}{0.47,0.53,0.60}
\definecolor{LightSlateGrey}{rgb}{0.47,0.53,0.60}
\definecolor{LightSteelBlue1}{rgb}{0.79,0.88,1.00}
\definecolor{LightSteelBlue2}{rgb}{0.74,0.82,0.93}
\definecolor{LightSteelBlue3}{rgb}{0.64,0.71,0.80}
\definecolor{LightSteelBlue4}{rgb}{0.43,0.48,0.55}
\definecolor{LightSteelBlue}{rgb}{0.69,0.77,0.87}
\definecolor{LightYellow1}{rgb}{1.00,1.00,0.88}
\definecolor{LightYellow2}{rgb}{0.93,0.93,0.82}
\definecolor{LightYellow3}{rgb}{0.80,0.80,0.71}
\definecolor{LightYellow4}{rgb}{0.55,0.55,0.48}
\definecolor{LightYellow}{rgb}{1.00,1.00,0.88}
\definecolor{LimeGreen}{rgb}{0.20,0.80,0.20}
\definecolor{MediumAquamarine}{rgb}{0.40,0.80,0.67}
\definecolor{MediumBlue}{rgb}{0.00,0.00,0.80}
\definecolor{MediumOrchid1}{rgb}{0.88,0.40,1.00}
\definecolor{MediumOrchid2}{rgb}{0.82,0.37,0.93}
\definecolor{MediumOrchid3}{rgb}{0.71,0.32,0.80}
\definecolor{MediumOrchid4}{rgb}{0.48,0.22,0.55}
\definecolor{MediumOrchid}{rgb}{0.73,0.33,0.83}
\definecolor{MediumPurple1}{rgb}{0.67,0.51,1.00}
\definecolor{MediumPurple2}{rgb}{0.62,0.47,0.93}
\definecolor{MediumPurple3}{rgb}{0.54,0.41,0.80}
\definecolor{MediumPurple4}{rgb}{0.36,0.28,0.55}
\definecolor{MediumPurple}{rgb}{0.58,0.44,0.86}
\definecolor{MediumSeaGreen}{rgb}{0.24,0.70,0.44}
\definecolor{MediumSlateBlue}{rgb}{0.48,0.41,0.93}
\definecolor{MediumSpringGreen}{rgb}{0.00,0.98,0.60}
\definecolor{MediumTurquoise}{rgb}{0.28,0.82,0.80}
\definecolor{MediumVioletRed}{rgb}{0.78,0.08,0.52}
\definecolor{MidnightBlue}{rgb}{0.10,0.10,0.44}
\definecolor{MintCream}{rgb}{0.96,1.00,0.98}
\definecolor{MistyRose1}{rgb}{1.00,0.89,0.88}
\definecolor{MistyRose2}{rgb}{0.93,0.84,0.82}
\definecolor{MistyRose3}{rgb}{0.80,0.72,0.71}
\definecolor{MistyRose4}{rgb}{0.55,0.49,0.48}
\definecolor{MistyRose}{rgb}{1.00,0.89,0.88}
\definecolor{NavajoWhite1}{rgb}{1.00,0.87,0.68}
\definecolor{NavajoWhite2}{rgb}{0.93,0.81,0.63}
\definecolor{NavajoWhite3}{rgb}{0.80,0.70,0.55}
\definecolor{NavajoWhite4}{rgb}{0.55,0.47,0.37}
\definecolor{NavajoWhite}{rgb}{1.00,0.87,0.68}
\definecolor{NavyBlue}{rgb}{0.00,0.00,0.50}
\definecolor{OldLace}{rgb}{0.99,0.96,0.90}
\definecolor{OliveDrab1}{rgb}{0.75,1.00,0.24}
\definecolor{OliveDrab2}{rgb}{0.70,0.93,0.23}
\definecolor{OliveDrab3}{rgb}{0.60,0.80,0.20}
\definecolor{OliveDrab4}{rgb}{0.41,0.55,0.13}
\definecolor{OliveDrab}{rgb}{0.42,0.56,0.14}
\definecolor{OrangeRed1}{rgb}{1.00,0.27,0.00}
\definecolor{OrangeRed2}{rgb}{0.93,0.25,0.00}
\definecolor{OrangeRed3}{rgb}{0.80,0.22,0.00}
\definecolor{OrangeRed4}{rgb}{0.55,0.15,0.00}
\definecolor{OrangeRed}{rgb}{1.00,0.27,0.00}
\definecolor{PaleGoldenrod}{rgb}{0.93,0.91,0.67}
\definecolor{PaleGreen1}{rgb}{0.60,1.00,0.60}
\definecolor{PaleGreen2}{rgb}{0.56,0.93,0.56}
\definecolor{PaleGreen3}{rgb}{0.49,0.80,0.49}
\definecolor{PaleGreen4}{rgb}{0.33,0.55,0.33}
\definecolor{PaleGreen}{rgb}{0.60,0.98,0.60}
\definecolor{PaleTurquoise1}{rgb}{0.73,1.00,1.00}
\definecolor{PaleTurquoise2}{rgb}{0.68,0.93,0.93}
\definecolor{PaleTurquoise3}{rgb}{0.59,0.80,0.80}
\definecolor{PaleTurquoise4}{rgb}{0.40,0.55,0.55}
\definecolor{PaleTurquoise}{rgb}{0.69,0.93,0.93}
\definecolor{PaleVioletRed1}{rgb}{1.00,0.51,0.67}
\definecolor{PaleVioletRed2}{rgb}{0.93,0.47,0.62}
\definecolor{PaleVioletRed3}{rgb}{0.80,0.41,0.54}
\definecolor{PaleVioletRed4}{rgb}{0.55,0.28,0.36}
\definecolor{PaleVioletRed}{rgb}{0.86,0.44,0.58}
\definecolor{PapayaWhip}{rgb}{1.00,0.94,0.84}
\definecolor{PeachPuff1}{rgb}{1.00,0.85,0.73}
\definecolor{PeachPuff2}{rgb}{0.93,0.80,0.68}
\definecolor{PeachPuff3}{rgb}{0.80,0.69,0.58}
\definecolor{PeachPuff4}{rgb}{0.55,0.47,0.40}
\definecolor{PeachPuff}{rgb}{1.00,0.85,0.73}
\definecolor{PowderBlue}{rgb}{0.69,0.88,0.90}
\definecolor{RosyBrown1}{rgb}{1.00,0.76,0.76}
\definecolor{RosyBrown2}{rgb}{0.93,0.71,0.71}
\definecolor{RosyBrown3}{rgb}{0.80,0.61,0.61}
\definecolor{RosyBrown4}{rgb}{0.55,0.41,0.41}
\definecolor{RosyBrown}{rgb}{0.74,0.56,0.56}
\definecolor{RoyalBlue1}{rgb}{0.28,0.46,1.00}
\definecolor{RoyalBlue2}{rgb}{0.26,0.43,0.93}
\definecolor{RoyalBlue3}{rgb}{0.23,0.37,0.80}
\definecolor{RoyalBlue4}{rgb}{0.15,0.25,0.55}
\definecolor{RoyalBlue}{rgb}{0.25,0.41,0.88}
\definecolor{SaddleBrown}{rgb}{0.55,0.27,0.07}
\definecolor{SandyBrown}{rgb}{0.96,0.64,0.38}
\definecolor{SeaGreen1}{rgb}{0.33,1.00,0.62}
\definecolor{SeaGreen2}{rgb}{0.31,0.93,0.58}
\definecolor{SeaGreen3}{rgb}{0.26,0.80,0.50}
\definecolor{SeaGreen4}{rgb}{0.18,0.55,0.34}
\definecolor{SeaGreen}{rgb}{0.18,0.55,0.34}
\definecolor{SkyBlue1}{rgb}{0.53,0.81,1.00}
\definecolor{SkyBlue2}{rgb}{0.49,0.75,0.93}
\definecolor{SkyBlue3}{rgb}{0.42,0.65,0.80}
\definecolor{SkyBlue4}{rgb}{0.29,0.44,0.55}
\definecolor{SkyBlue}{rgb}{0.53,0.81,0.92}
\definecolor{SlateBlue1}{rgb}{0.51,0.44,1.00}
\definecolor{SlateBlue2}{rgb}{0.48,0.40,0.93}
\definecolor{SlateBlue3}{rgb}{0.41,0.35,0.80}
\definecolor{SlateBlue4}{rgb}{0.28,0.24,0.55}
\definecolor{SlateBlue}{rgb}{0.42,0.35,0.80}
\definecolor{SlateGray1}{rgb}{0.78,0.89,1.00}
\definecolor{SlateGray2}{rgb}{0.73,0.83,0.93}
\definecolor{SlateGray3}{rgb}{0.62,0.71,0.80}
\definecolor{SlateGray4}{rgb}{0.42,0.48,0.55}
\definecolor{SlateGray}{rgb}{0.44,0.50,0.56}
\definecolor{SlateGrey}{rgb}{0.44,0.50,0.56}
\definecolor{SpringGreen1}{rgb}{0.00,1.00,0.50}
\definecolor{SpringGreen2}{rgb}{0.00,0.93,0.46}
\definecolor{SpringGreen3}{rgb}{0.00,0.80,0.40}
\definecolor{SpringGreen4}{rgb}{0.00,0.55,0.27}
\definecolor{SpringGreen}{rgb}{0.00,1.00,0.50}
\definecolor{SteelBlue1}{rgb}{0.39,0.72,1.00}
\definecolor{SteelBlue2}{rgb}{0.36,0.67,0.93}
\definecolor{SteelBlue3}{rgb}{0.31,0.58,0.80}
\definecolor{SteelBlue4}{rgb}{0.21,0.39,0.55}
\definecolor{SteelBlue}{rgb}{0.27,0.51,0.71}
\definecolor{VioletRed1}{rgb}{1.00,0.24,0.59}
\definecolor{VioletRed2}{rgb}{0.93,0.23,0.55}
\definecolor{VioletRed3}{rgb}{0.80,0.20,0.47}
\definecolor{VioletRed4}{rgb}{0.55,0.13,0.32}
\definecolor{VioletRed}{rgb}{0.82,0.13,0.56}
\definecolor{WhiteSmoke}{rgb}{0.96,0.96,0.96}
\definecolor{YellowGreen}{rgb}{0.60,0.80,0.20}
\definecolor{aliceblue}{rgb}{0.94,0.97,1.00}
\definecolor{antiquewhite}{rgb}{0.98,0.92,0.84}
\definecolor{aquamarine1}{rgb}{0.50,1.00,0.83}
\definecolor{aquamarine2}{rgb}{0.46,0.93,0.78}
\definecolor{aquamarine3}{rgb}{0.40,0.80,0.67}
\definecolor{aquamarine4}{rgb}{0.27,0.55,0.45}
\definecolor{aquamarine}{rgb}{0.50,1.00,0.83}
\definecolor{azure1}{rgb}{0.94,1.00,1.00}
\definecolor{azure2}{rgb}{0.88,0.93,0.93}
\definecolor{azure3}{rgb}{0.76,0.80,0.80}
\definecolor{azure4}{rgb}{0.51,0.55,0.55}
\definecolor{azure}{rgb}{0.94,1.00,1.00}
\definecolor{beige}{rgb}{0.96,0.96,0.86}
\definecolor{bisque1}{rgb}{1.00,0.89,0.77}
\definecolor{bisque2}{rgb}{0.93,0.84,0.72}
\definecolor{bisque3}{rgb}{0.80,0.72,0.62}
\definecolor{bisque4}{rgb}{0.55,0.49,0.42}
\definecolor{bisque}{rgb}{1.00,0.89,0.77}
\definecolor{black}{rgb}{0.00,0.00,0.00}
\definecolor{blanchedalmond}{rgb}{1.00,0.92,0.80}
\definecolor{blue1}{rgb}{0.00,0.00,1.00}
\definecolor{blue2}{rgb}{0.00,0.00,0.93}
\definecolor{blue3}{rgb}{0.00,0.00,0.80}
\definecolor{blue4}{rgb}{0.00,0.00,0.55}
\definecolor{blueviolet}{rgb}{0.54,0.17,0.89}
\definecolor{blue}{rgb}{0.00,0.00,1.00}
\definecolor{brown1}{rgb}{1.00,0.25,0.25}
\definecolor{brown2}{rgb}{0.93,0.23,0.23}
\definecolor{brown3}{rgb}{0.80,0.20,0.20}
\definecolor{brown4}{rgb}{0.55,0.14,0.14}
\definecolor{brown}{rgb}{0.65,0.16,0.16}
\definecolor{burlywood1}{rgb}{1.00,0.83,0.61}
\definecolor{burlywood2}{rgb}{0.93,0.77,0.57}
\definecolor{burlywood3}{rgb}{0.80,0.67,0.49}
\definecolor{burlywood4}{rgb}{0.55,0.45,0.33}
\definecolor{burlywood}{rgb}{0.87,0.72,0.53}
\definecolor{cadetblue}{rgb}{0.37,0.62,0.63}
\definecolor{chartreuse1}{rgb}{0.50,1.00,0.00}
\definecolor{chartreuse2}{rgb}{0.46,0.93,0.00}
\definecolor{chartreuse3}{rgb}{0.40,0.80,0.00}
\definecolor{chartreuse4}{rgb}{0.27,0.55,0.00}
\definecolor{chartreuse}{rgb}{0.50,1.00,0.00}
\definecolor{chocolate1}{rgb}{1.00,0.50,0.14}
\definecolor{chocolate2}{rgb}{0.93,0.46,0.13}
\definecolor{chocolate3}{rgb}{0.80,0.40,0.11}
\definecolor{chocolate4}{rgb}{0.55,0.27,0.07}
\definecolor{chocolate}{rgb}{0.82,0.41,0.12}
\definecolor{coral1}{rgb}{1.00,0.45,0.34}
\definecolor{coral2}{rgb}{0.93,0.42,0.31}
\definecolor{coral3}{rgb}{0.80,0.36,0.27}
\definecolor{coral4}{rgb}{0.55,0.24,0.18}
\definecolor{coral}{rgb}{1.00,0.50,0.31}
\definecolor{cornflowerblue}{rgb}{0.39,0.58,0.93}
\definecolor{cornsilk1}{rgb}{1.00,0.97,0.86}
\definecolor{cornsilk2}{rgb}{0.93,0.91,0.80}
\definecolor{cornsilk3}{rgb}{0.80,0.78,0.69}
\definecolor{cornsilk4}{rgb}{0.55,0.53,0.47}
\definecolor{cornsilk}{rgb}{1.00,0.97,0.86}
\definecolor{cyan1}{rgb}{0.00,1.00,1.00}
\definecolor{cyan2}{rgb}{0.00,0.93,0.93}
\definecolor{cyan3}{rgb}{0.00,0.80,0.80}
\definecolor{cyan4}{rgb}{0.00,0.55,0.55}
\definecolor{cyan}{rgb}{0.00,1.00,1.00}
\definecolor{darkblue}{rgb}{0.00,0.00,0.55}
\definecolor{darkcyan}{rgb}{0.00,0.55,0.55}
\definecolor{darkgoldenrod}{rgb}{0.72,0.53,0.04}
\definecolor{darkgray}{rgb}{0.66,0.66,0.66}
\definecolor{darkgreen}{rgb}{0.00,0.39,0.00}
\definecolor{darkgrey}{rgb}{0.66,0.66,0.66}
\definecolor{darkkhaki}{rgb}{0.74,0.72,0.42}
\definecolor{darkmagenta}{rgb}{0.55,0.00,0.55}
\definecolor{darkolive}{rgb}{0.33,0.42,0.18}
\definecolor{darkorange}{rgb}{1.00,0.55,0.00}
\definecolor{darkorchid}{rgb}{0.60,0.20,0.80}
\definecolor{darkred}{rgb}{0.55,0.00,0.00}
\definecolor{darksalmon}{rgb}{0.91,0.59,0.48}
\definecolor{darksea}{rgb}{0.56,0.74,0.56}
\definecolor{darkslate}{rgb}{0.18,0.31,0.31}
\definecolor{darkslate}{rgb}{0.18,0.31,0.31}
\definecolor{darkslate}{rgb}{0.28,0.24,0.55}
\definecolor{darkturquoise}{rgb}{0.00,0.81,0.82}
\definecolor{darkviolet}{rgb}{0.58,0.00,0.83}
\definecolor{deeppink}{rgb}{1.00,0.08,0.58}
\definecolor{deepsky}{rgb}{0.00,0.75,1.00}
\definecolor{dimgray}{rgb}{0.41,0.41,0.41}
\definecolor{dimgrey}{rgb}{0.41,0.41,0.41}
\definecolor{dodgerblue}{rgb}{0.12,0.56,1.00}
\definecolor{firebrick1}{rgb}{1.00,0.19,0.19}
\definecolor{firebrick2}{rgb}{0.93,0.17,0.17}
\definecolor{firebrick3}{rgb}{0.80,0.15,0.15}
\definecolor{firebrick4}{rgb}{0.55,0.10,0.10}
\definecolor{firebrick}{rgb}{0.70,0.13,0.13}
\definecolor{floralwhite}{rgb}{1.00,0.98,0.94}
\definecolor{forestgreen}{rgb}{0.13,0.55,0.13}
\definecolor{gainsboro}{rgb}{0.86,0.86,0.86}
\definecolor{ghostwhite}{rgb}{0.97,0.97,1.00}
\definecolor{gold1}{rgb}{1.00,0.84,0.00}
\definecolor{gold2}{rgb}{0.93,0.79,0.00}
\definecolor{gold3}{rgb}{0.80,0.68,0.00}
\definecolor{gold4}{rgb}{0.55,0.46,0.00}
\definecolor{goldenrod1}{rgb}{1.00,0.76,0.15}
\definecolor{goldenrod2}{rgb}{0.93,0.71,0.13}
\definecolor{goldenrod3}{rgb}{0.80,0.61,0.11}
\definecolor{goldenrod4}{rgb}{0.55,0.41,0.08}
\definecolor{goldenrod}{rgb}{0.85,0.65,0.13}
\definecolor{gold}{rgb}{1.00,0.84,0.00}
\definecolor{gray0}{rgb}{0.00,0.00,0.00}
\definecolor{gray100}{rgb}{1.00,1.00,1.00}
\definecolor{gray10}{rgb}{0.10,0.10,0.10}
\definecolor{gray11}{rgb}{0.11,0.11,0.11}
\definecolor{gray12}{rgb}{0.12,0.12,0.12}
\definecolor{gray13}{rgb}{0.13,0.13,0.13}
\definecolor{gray14}{rgb}{0.14,0.14,0.14}
\definecolor{gray15}{rgb}{0.15,0.15,0.15}
\definecolor{gray16}{rgb}{0.16,0.16,0.16}
\definecolor{gray17}{rgb}{0.17,0.17,0.17}
\definecolor{gray18}{rgb}{0.18,0.18,0.18}
\definecolor{gray19}{rgb}{0.19,0.19,0.19}
\definecolor{gray1}{rgb}{0.01,0.01,0.01}
\definecolor{gray20}{rgb}{0.20,0.20,0.20}
\definecolor{gray21}{rgb}{0.21,0.21,0.21}
\definecolor{gray22}{rgb}{0.22,0.22,0.22}
\definecolor{gray23}{rgb}{0.23,0.23,0.23}
\definecolor{gray24}{rgb}{0.24,0.24,0.24}
\definecolor{gray25}{rgb}{0.25,0.25,0.25}
\definecolor{gray26}{rgb}{0.26,0.26,0.26}
\definecolor{gray27}{rgb}{0.27,0.27,0.27}
\definecolor{gray28}{rgb}{0.28,0.28,0.28}
\definecolor{gray29}{rgb}{0.29,0.29,0.29}
\definecolor{gray2}{rgb}{0.02,0.02,0.02}
\definecolor{gray30}{rgb}{0.30,0.30,0.30}
\definecolor{gray31}{rgb}{0.31,0.31,0.31}
\definecolor{gray32}{rgb}{0.32,0.32,0.32}
\definecolor{gray33}{rgb}{0.33,0.33,0.33}
\definecolor{gray34}{rgb}{0.34,0.34,0.34}
\definecolor{gray35}{rgb}{0.35,0.35,0.35}
\definecolor{gray36}{rgb}{0.36,0.36,0.36}
\definecolor{gray37}{rgb}{0.37,0.37,0.37}
\definecolor{gray38}{rgb}{0.38,0.38,0.38}
\definecolor{gray39}{rgb}{0.39,0.39,0.39}
\definecolor{gray3}{rgb}{0.03,0.03,0.03}
\definecolor{gray40}{rgb}{0.40,0.40,0.40}
\definecolor{gray41}{rgb}{0.41,0.41,0.41}
\definecolor{gray42}{rgb}{0.42,0.42,0.42}
\definecolor{gray43}{rgb}{0.43,0.43,0.43}
\definecolor{gray44}{rgb}{0.44,0.44,0.44}
\definecolor{gray45}{rgb}{0.45,0.45,0.45}
\definecolor{gray46}{rgb}{0.46,0.46,0.46}
\definecolor{gray47}{rgb}{0.47,0.47,0.47}
\definecolor{gray48}{rgb}{0.48,0.48,0.48}
\definecolor{gray49}{rgb}{0.49,0.49,0.49}
\definecolor{gray4}{rgb}{0.04,0.04,0.04}
\definecolor{gray50}{rgb}{0.50,0.50,0.50}
\definecolor{gray51}{rgb}{0.51,0.51,0.51}
\definecolor{gray52}{rgb}{0.52,0.52,0.52}
\definecolor{gray53}{rgb}{0.53,0.53,0.53}
\definecolor{gray54}{rgb}{0.54,0.54,0.54}
\definecolor{gray55}{rgb}{0.55,0.55,0.55}
\definecolor{gray56}{rgb}{0.56,0.56,0.56}
\definecolor{gray57}{rgb}{0.57,0.57,0.57}
\definecolor{gray58}{rgb}{0.58,0.58,0.58}
\definecolor{gray59}{rgb}{0.59,0.59,0.59}
\definecolor{gray5}{rgb}{0.05,0.05,0.05}
\definecolor{gray60}{rgb}{0.60,0.60,0.60}
\definecolor{gray61}{rgb}{0.61,0.61,0.61}
\definecolor{gray62}{rgb}{0.62,0.62,0.62}
\definecolor{gray63}{rgb}{0.63,0.63,0.63}
\definecolor{gray64}{rgb}{0.64,0.64,0.64}
\definecolor{gray65}{rgb}{0.65,0.65,0.65}
\definecolor{gray66}{rgb}{0.66,0.66,0.66}
\definecolor{gray67}{rgb}{0.67,0.67,0.67}
\definecolor{gray68}{rgb}{0.68,0.68,0.68}
\definecolor{gray69}{rgb}{0.69,0.69,0.69}
\definecolor{gray6}{rgb}{0.06,0.06,0.06}
\definecolor{gray70}{rgb}{0.70,0.70,0.70}
\definecolor{gray71}{rgb}{0.71,0.71,0.71}
\definecolor{gray72}{rgb}{0.72,0.72,0.72}
\definecolor{gray73}{rgb}{0.73,0.73,0.73}
\definecolor{gray74}{rgb}{0.74,0.74,0.74}
\definecolor{gray75}{rgb}{0.75,0.75,0.75}
\definecolor{gray76}{rgb}{0.76,0.76,0.76}
\definecolor{gray77}{rgb}{0.77,0.77,0.77}
\definecolor{gray78}{rgb}{0.78,0.78,0.78}
\definecolor{gray79}{rgb}{0.79,0.79,0.79}
\definecolor{gray7}{rgb}{0.07,0.07,0.07}
\definecolor{gray80}{rgb}{0.80,0.80,0.80}
\definecolor{gray81}{rgb}{0.81,0.81,0.81}
\definecolor{gray82}{rgb}{0.82,0.82,0.82}
\definecolor{gray83}{rgb}{0.83,0.83,0.83}
\definecolor{gray84}{rgb}{0.84,0.84,0.84}
\definecolor{gray85}{rgb}{0.85,0.85,0.85}
\definecolor{gray86}{rgb}{0.86,0.86,0.86}
\definecolor{gray87}{rgb}{0.87,0.87,0.87}
\definecolor{gray88}{rgb}{0.88,0.88,0.88}
\definecolor{gray89}{rgb}{0.89,0.89,0.89}
\definecolor{gray8}{rgb}{0.08,0.08,0.08}
\definecolor{gray90}{rgb}{0.90,0.90,0.90}
\definecolor{gray91}{rgb}{0.91,0.91,0.91}
\definecolor{gray92}{rgb}{0.92,0.92,0.92}
\definecolor{gray93}{rgb}{0.93,0.93,0.93}
\definecolor{gray94}{rgb}{0.94,0.94,0.94}
\definecolor{gray95}{rgb}{0.95,0.95,0.95}
\definecolor{gray96}{rgb}{0.96,0.96,0.96}
\definecolor{gray97}{rgb}{0.97,0.97,0.97}
\definecolor{gray98}{rgb}{0.98,0.98,0.98}
\definecolor{gray99}{rgb}{0.99,0.99,0.99}
\definecolor{gray9}{rgb}{0.09,0.09,0.09}
\definecolor{gray}{rgb}{0.75,0.75,0.75}
\definecolor{green1}{rgb}{0.00,1.00,0.00}
\definecolor{green2}{rgb}{0.00,0.93,0.00}
\definecolor{green3}{rgb}{0.00,0.80,0.00}
\definecolor{green4}{rgb}{0.00,0.55,0.00}
\definecolor{greenyellow}{rgb}{0.68,1.00,0.18}
\definecolor{green}{rgb}{0.00,1.00,0.00}
\definecolor{grey0}{rgb}{0.00,0.00,0.00}
\definecolor{grey100}{rgb}{1.00,1.00,1.00}
\definecolor{grey10}{rgb}{0.10,0.10,0.10}
\definecolor{grey11}{rgb}{0.11,0.11,0.11}
\definecolor{grey12}{rgb}{0.12,0.12,0.12}
\definecolor{grey13}{rgb}{0.13,0.13,0.13}
\definecolor{grey14}{rgb}{0.14,0.14,0.14}
\definecolor{grey15}{rgb}{0.15,0.15,0.15}
\definecolor{grey16}{rgb}{0.16,0.16,0.16}
\definecolor{grey17}{rgb}{0.17,0.17,0.17}
\definecolor{grey18}{rgb}{0.18,0.18,0.18}
\definecolor{grey19}{rgb}{0.19,0.19,0.19}
\definecolor{grey1}{rgb}{0.01,0.01,0.01}
\definecolor{grey20}{rgb}{0.20,0.20,0.20}
\definecolor{grey21}{rgb}{0.21,0.21,0.21}
\definecolor{grey22}{rgb}{0.22,0.22,0.22}
\definecolor{grey23}{rgb}{0.23,0.23,0.23}
\definecolor{grey24}{rgb}{0.24,0.24,0.24}
\definecolor{grey25}{rgb}{0.25,0.25,0.25}
\definecolor{grey26}{rgb}{0.26,0.26,0.26}
\definecolor{grey27}{rgb}{0.27,0.27,0.27}
\definecolor{grey28}{rgb}{0.28,0.28,0.28}
\definecolor{grey29}{rgb}{0.29,0.29,0.29}
\definecolor{grey2}{rgb}{0.02,0.02,0.02}
\definecolor{grey30}{rgb}{0.30,0.30,0.30}
\definecolor{grey31}{rgb}{0.31,0.31,0.31}
\definecolor{grey32}{rgb}{0.32,0.32,0.32}
\definecolor{grey33}{rgb}{0.33,0.33,0.33}
\definecolor{grey34}{rgb}{0.34,0.34,0.34}
\definecolor{grey35}{rgb}{0.35,0.35,0.35}
\definecolor{grey36}{rgb}{0.36,0.36,0.36}
\definecolor{grey37}{rgb}{0.37,0.37,0.37}
\definecolor{grey38}{rgb}{0.38,0.38,0.38}
\definecolor{grey39}{rgb}{0.39,0.39,0.39}
\definecolor{grey3}{rgb}{0.03,0.03,0.03}
\definecolor{grey40}{rgb}{0.40,0.40,0.40}
\definecolor{grey41}{rgb}{0.41,0.41,0.41}
\definecolor{grey42}{rgb}{0.42,0.42,0.42}
\definecolor{grey43}{rgb}{0.43,0.43,0.43}
\definecolor{grey44}{rgb}{0.44,0.44,0.44}
\definecolor{grey45}{rgb}{0.45,0.45,0.45}
\definecolor{grey46}{rgb}{0.46,0.46,0.46}
\definecolor{grey47}{rgb}{0.47,0.47,0.47}
\definecolor{grey48}{rgb}{0.48,0.48,0.48}
\definecolor{grey49}{rgb}{0.49,0.49,0.49}
\definecolor{grey4}{rgb}{0.04,0.04,0.04}
\definecolor{grey50}{rgb}{0.50,0.50,0.50}
\definecolor{grey51}{rgb}{0.51,0.51,0.51}
\definecolor{grey52}{rgb}{0.52,0.52,0.52}
\definecolor{grey53}{rgb}{0.53,0.53,0.53}
\definecolor{grey54}{rgb}{0.54,0.54,0.54}
\definecolor{grey55}{rgb}{0.55,0.55,0.55}
\definecolor{grey56}{rgb}{0.56,0.56,0.56}
\definecolor{grey57}{rgb}{0.57,0.57,0.57}
\definecolor{grey58}{rgb}{0.58,0.58,0.58}
\definecolor{grey59}{rgb}{0.59,0.59,0.59}
\definecolor{grey5}{rgb}{0.05,0.05,0.05}
\definecolor{grey60}{rgb}{0.60,0.60,0.60}
\definecolor{grey61}{rgb}{0.61,0.61,0.61}
\definecolor{grey62}{rgb}{0.62,0.62,0.62}
\definecolor{grey63}{rgb}{0.63,0.63,0.63}
\definecolor{grey64}{rgb}{0.64,0.64,0.64}
\definecolor{grey65}{rgb}{0.65,0.65,0.65}
\definecolor{grey66}{rgb}{0.66,0.66,0.66}
\definecolor{grey67}{rgb}{0.67,0.67,0.67}
\definecolor{grey68}{rgb}{0.68,0.68,0.68}
\definecolor{grey69}{rgb}{0.69,0.69,0.69}
\definecolor{grey6}{rgb}{0.06,0.06,0.06}
\definecolor{grey70}{rgb}{0.70,0.70,0.70}
\definecolor{grey71}{rgb}{0.71,0.71,0.71}
\definecolor{grey72}{rgb}{0.72,0.72,0.72}
\definecolor{grey73}{rgb}{0.73,0.73,0.73}
\definecolor{grey74}{rgb}{0.74,0.74,0.74}
\definecolor{grey75}{rgb}{0.75,0.75,0.75}
\definecolor{grey76}{rgb}{0.76,0.76,0.76}
\definecolor{grey77}{rgb}{0.77,0.77,0.77}
\definecolor{grey78}{rgb}{0.78,0.78,0.78}
\definecolor{grey79}{rgb}{0.79,0.79,0.79}
\definecolor{grey7}{rgb}{0.07,0.07,0.07}
\definecolor{grey80}{rgb}{0.80,0.80,0.80}
\definecolor{grey81}{rgb}{0.81,0.81,0.81}
\definecolor{grey82}{rgb}{0.82,0.82,0.82}
\definecolor{grey83}{rgb}{0.83,0.83,0.83}
\definecolor{grey84}{rgb}{0.84,0.84,0.84}
\definecolor{grey85}{rgb}{0.85,0.85,0.85}
\definecolor{grey86}{rgb}{0.86,0.86,0.86}
\definecolor{grey87}{rgb}{0.87,0.87,0.87}
\definecolor{grey88}{rgb}{0.88,0.88,0.88}
\definecolor{grey89}{rgb}{0.89,0.89,0.89}
\definecolor{grey8}{rgb}{0.08,0.08,0.08}
\definecolor{grey90}{rgb}{0.90,0.90,0.90}
\definecolor{grey91}{rgb}{0.91,0.91,0.91}
\definecolor{grey92}{rgb}{0.92,0.92,0.92}
\definecolor{grey93}{rgb}{0.93,0.93,0.93}
\definecolor{grey94}{rgb}{0.94,0.94,0.94}
\definecolor{grey95}{rgb}{0.95,0.95,0.95}
\definecolor{grey96}{rgb}{0.96,0.96,0.96}
\definecolor{grey97}{rgb}{0.97,0.97,0.97}
\definecolor{grey98}{rgb}{0.98,0.98,0.98}
\definecolor{grey99}{rgb}{0.99,0.99,0.99}
\definecolor{grey9}{rgb}{0.09,0.09,0.09}
\definecolor{grey}{rgb}{0.75,0.75,0.75}
\definecolor{honeydew1}{rgb}{0.94,1.00,0.94}
\definecolor{honeydew2}{rgb}{0.88,0.93,0.88}
\definecolor{honeydew3}{rgb}{0.76,0.80,0.76}
\definecolor{honeydew4}{rgb}{0.51,0.55,0.51}
\definecolor{honeydew}{rgb}{0.94,1.00,0.94}
\definecolor{hotpink}{rgb}{1.00,0.41,0.71}
\definecolor{indianred}{rgb}{0.80,0.36,0.36}
\definecolor{ivory1}{rgb}{1.00,1.00,0.94}
\definecolor{ivory2}{rgb}{0.93,0.93,0.88}
\definecolor{ivory3}{rgb}{0.80,0.80,0.76}
\definecolor{ivory4}{rgb}{0.55,0.55,0.51}
\definecolor{ivory}{rgb}{1.00,1.00,0.94}
\definecolor{khaki1}{rgb}{1.00,0.96,0.56}
\definecolor{khaki2}{rgb}{0.93,0.90,0.52}
\definecolor{khaki3}{rgb}{0.80,0.78,0.45}
\definecolor{khaki4}{rgb}{0.55,0.53,0.31}
\definecolor{khaki}{rgb}{0.94,0.90,0.55}
\definecolor{lavenderblush}{rgb}{1.00,0.94,0.96}
\definecolor{lavender}{rgb}{0.90,0.90,0.98}
\definecolor{lawngreen}{rgb}{0.49,0.99,0.00}
\definecolor{lemonchiffon}{rgb}{1.00,0.98,0.80}
\definecolor{lightblue}{rgb}{0.68,0.85,0.90}
\definecolor{lightcoral}{rgb}{0.94,0.50,0.50}
\definecolor{lightcyan}{rgb}{0.88,1.00,1.00}
\definecolor{lightgoldenrod}{rgb}{0.93,0.87,0.51}
\definecolor{lightgoldenrod}{rgb}{0.98,0.98,0.82}
\definecolor{lightgray}{rgb}{0.83,0.83,0.83}
\definecolor{lightgreen}{rgb}{0.56,0.93,0.56}
\definecolor{lightgrey}{rgb}{0.83,0.83,0.83}
\definecolor{lightpink}{rgb}{1.00,0.71,0.76}
\definecolor{lightsalmon}{rgb}{1.00,0.63,0.48}
\definecolor{lightsea}{rgb}{0.13,0.70,0.67}
\definecolor{lightsky}{rgb}{0.53,0.81,0.98}
\definecolor{lightslate}{rgb}{0.47,0.53,0.60}
\definecolor{lightslate}{rgb}{0.47,0.53,0.60}
\definecolor{lightslate}{rgb}{0.52,0.44,1.00}
\definecolor{lightsteel}{rgb}{0.69,0.77,0.87}
\definecolor{lightyellow}{rgb}{1.00,1.00,0.88}
\definecolor{limegreen}{rgb}{0.20,0.80,0.20}
\definecolor{linen}{rgb}{0.98,0.94,0.90}
\definecolor{magenta1}{rgb}{1.00,0.00,1.00}
\definecolor{magenta2}{rgb}{0.93,0.00,0.93}
\definecolor{magenta3}{rgb}{0.80,0.00,0.80}
\definecolor{magenta4}{rgb}{0.55,0.00,0.55}
\definecolor{magenta}{rgb}{1.00,0.00,1.00}
\definecolor{maroon1}{rgb}{1.00,0.20,0.70}
\definecolor{maroon2}{rgb}{0.93,0.19,0.65}
\definecolor{maroon3}{rgb}{0.80,0.16,0.56}
\definecolor{maroon4}{rgb}{0.55,0.11,0.38}
\definecolor{maroon}{rgb}{0.69,0.19,0.38}
\definecolor{mediumaquamarine}{rgb}{0.40,0.80,0.67}
\definecolor{mediumblue}{rgb}{0.00,0.00,0.80}
\definecolor{mediumorchid}{rgb}{0.73,0.33,0.83}
\definecolor{mediumpurple}{rgb}{0.58,0.44,0.86}
\definecolor{mediumsea}{rgb}{0.24,0.70,0.44}
\definecolor{mediumslate}{rgb}{0.48,0.41,0.93}
\definecolor{mediumspring}{rgb}{0.00,0.98,0.60}
\definecolor{mediumturquoise}{rgb}{0.28,0.82,0.80}
\definecolor{mediumviolet}{rgb}{0.78,0.08,0.52}
\definecolor{midnightblue}{rgb}{0.10,0.10,0.44}
\definecolor{mintcream}{rgb}{0.96,1.00,0.98}
\definecolor{mistyrose}{rgb}{1.00,0.89,0.88}
\definecolor{moccasin}{rgb}{1.00,0.89,0.71}
\definecolor{navajowhite}{rgb}{1.00,0.87,0.68}
\definecolor{navyblue}{rgb}{0.00,0.00,0.50}
\definecolor{navy}{rgb}{0.00,0.00,0.50}
\definecolor{oldlace}{rgb}{0.99,0.96,0.90}
\definecolor{olivedrab}{rgb}{0.42,0.56,0.14}
\definecolor{orange1}{rgb}{1.00,0.65,0.00}
\definecolor{orange2}{rgb}{0.93,0.60,0.00}
\definecolor{orange3}{rgb}{0.80,0.52,0.00}
\definecolor{orange4}{rgb}{0.55,0.35,0.00}
\definecolor{orangered}{rgb}{1.00,0.27,0.00}
\definecolor{orange}{rgb}{1.00,0.65,0.00}
\definecolor{orchid1}{rgb}{1.00,0.51,0.98}
\definecolor{orchid2}{rgb}{0.93,0.48,0.91}
\definecolor{orchid3}{rgb}{0.80,0.41,0.79}
\definecolor{orchid4}{rgb}{0.55,0.28,0.54}
\definecolor{orchid}{rgb}{0.85,0.44,0.84}
\definecolor{palegoldenrod}{rgb}{0.93,0.91,0.67}
\definecolor{palegreen}{rgb}{0.60,0.98,0.60}
\definecolor{paleturquoise}{rgb}{0.69,0.93,0.93}
\definecolor{paleviolet}{rgb}{0.86,0.44,0.58}
\definecolor{papayawhip}{rgb}{1.00,0.94,0.84}
\definecolor{peachpuff}{rgb}{1.00,0.85,0.73}
\definecolor{peru}{rgb}{0.80,0.52,0.25}
\definecolor{pink1}{rgb}{1.00,0.71,0.77}
\definecolor{pink2}{rgb}{0.93,0.66,0.72}
\definecolor{pink3}{rgb}{0.80,0.57,0.62}
\definecolor{pink4}{rgb}{0.55,0.39,0.42}
\definecolor{pink}{rgb}{1.00,0.75,0.80}
\definecolor{plum1}{rgb}{1.00,0.73,1.00}
\definecolor{plum2}{rgb}{0.93,0.68,0.93}
\definecolor{plum3}{rgb}{0.80,0.59,0.80}
\definecolor{plum4}{rgb}{0.55,0.40,0.55}
\definecolor{plum}{rgb}{0.87,0.63,0.87}
\definecolor{powderblue}{rgb}{0.69,0.88,0.90}
\definecolor{purple1}{rgb}{0.61,0.19,1.00}
\definecolor{purple2}{rgb}{0.57,0.17,0.93}
\definecolor{purple3}{rgb}{0.49,0.15,0.80}
\definecolor{purple4}{rgb}{0.33,0.10,0.55}
\definecolor{purple}{rgb}{0.63,0.13,0.94}
\definecolor{red1}{rgb}{1.00,0.00,0.00}
\definecolor{red2}{rgb}{0.93,0.00,0.00}
\definecolor{red3}{rgb}{0.80,0.00,0.00}
\definecolor{red4}{rgb}{0.55,0.00,0.00}
\definecolor{red}{rgb}{1.00,0.00,0.00}
\definecolor{rosybrown}{rgb}{0.74,0.56,0.56}
\definecolor{royalblue}{rgb}{0.25,0.41,0.88}
\definecolor{saddlebrown}{rgb}{0.55,0.27,0.07}
\definecolor{salmon1}{rgb}{1.00,0.55,0.41}
\definecolor{salmon2}{rgb}{0.93,0.51,0.38}
\definecolor{salmon3}{rgb}{0.80,0.44,0.33}
\definecolor{salmon4}{rgb}{0.55,0.30,0.22}
\definecolor{salmon}{rgb}{0.98,0.50,0.45}
\definecolor{sandybrown}{rgb}{0.96,0.64,0.38}
\definecolor{seagreen}{rgb}{0.18,0.55,0.34}
\definecolor{seashell1}{rgb}{1.00,0.96,0.93}
\definecolor{seashell2}{rgb}{0.93,0.90,0.87}
\definecolor{seashell3}{rgb}{0.80,0.77,0.75}
\definecolor{seashell4}{rgb}{0.55,0.53,0.51}
\definecolor{seashell}{rgb}{1.00,0.96,0.93}
\definecolor{sienna1}{rgb}{1.00,0.51,0.28}
\definecolor{sienna2}{rgb}{0.93,0.47,0.26}
\definecolor{sienna3}{rgb}{0.80,0.41,0.22}
\definecolor{sienna4}{rgb}{0.55,0.28,0.15}
\definecolor{sienna}{rgb}{0.63,0.32,0.18}
\definecolor{skyblue}{rgb}{0.53,0.81,0.92}
\definecolor{slateblue}{rgb}{0.42,0.35,0.80}
\definecolor{slategray}{rgb}{0.44,0.50,0.56}
\definecolor{slategrey}{rgb}{0.44,0.50,0.56}
\definecolor{snow1}{rgb}{1.00,0.98,0.98}
\definecolor{snow2}{rgb}{0.93,0.91,0.91}
\definecolor{snow3}{rgb}{0.80,0.79,0.79}
\definecolor{snow4}{rgb}{0.55,0.54,0.54}
\definecolor{snow}{rgb}{1.00,0.98,0.98}
\definecolor{springgreen}{rgb}{0.00,1.00,0.50}
\definecolor{steelblue}{rgb}{0.27,0.51,0.71}
\definecolor{tan1}{rgb}{1.00,0.65,0.31}
\definecolor{tan2}{rgb}{0.93,0.60,0.29}
\definecolor{tan3}{rgb}{0.80,0.52,0.25}
\definecolor{tan4}{rgb}{0.55,0.35,0.17}
\definecolor{tan}{rgb}{0.82,0.71,0.55}
\definecolor{thistle1}{rgb}{1.00,0.88,1.00}
\definecolor{thistle2}{rgb}{0.93,0.82,0.93}
\definecolor{thistle3}{rgb}{0.80,0.71,0.80}
\definecolor{thistle4}{rgb}{0.55,0.48,0.55}
\definecolor{thistle}{rgb}{0.85,0.75,0.85}
\definecolor{tomato1}{rgb}{1.00,0.39,0.28}
\definecolor{tomato2}{rgb}{0.93,0.36,0.26}
\definecolor{tomato3}{rgb}{0.80,0.31,0.22}
\definecolor{tomato4}{rgb}{0.55,0.21,0.15}
\definecolor{tomato}{rgb}{1.00,0.39,0.28}
\definecolor{turquoise1}{rgb}{0.00,0.96,1.00}
\definecolor{turquoise2}{rgb}{0.00,0.90,0.93}
\definecolor{turquoise3}{rgb}{0.00,0.77,0.80}
\definecolor{turquoise4}{rgb}{0.00,0.53,0.55}
\definecolor{turquoise}{rgb}{0.25,0.88,0.82}
\definecolor{violetred}{rgb}{0.82,0.13,0.56}
\definecolor{violet}{rgb}{0.93,0.51,0.93}
\definecolor{wheat1}{rgb}{1.00,0.91,0.73}
\definecolor{wheat2}{rgb}{0.93,0.85,0.68}
\definecolor{wheat3}{rgb}{0.80,0.73,0.59}
\definecolor{wheat4}{rgb}{0.55,0.49,0.40}
\definecolor{wheat}{rgb}{0.96,0.87,0.70}
\definecolor{whitesmoke}{rgb}{0.96,0.96,0.96}
\definecolor{white}{rgb}{1.00,1.00,1.00}
\definecolor{yellow1}{rgb}{1.00,1.00,0.00}
\definecolor{yellow2}{rgb}{0.93,0.93,0.00}
\definecolor{yellow3}{rgb}{0.80,0.80,0.00}
\definecolor{yellow4}{rgb}{0.55,0.55,0.00}
\definecolor{yellowgreen}{rgb}{0.60,0.80,0.20}
\definecolor{yellow}{rgb}{1.00,1.00,0.00}

\bibpunct[; ]{(}{)}{;}{a}{}{;}

\title[Minor-merger driven star formation in early-type galaxies]
    {{\color{black}A coincidence of disturbed morphology and blue UV colour: minor-merger driven star
    formation in early-type galaxies at $z \sim 0.6$}}

\author[Sugata Kaviraj et al.]
{Sugata Kaviraj$^{1,2,3}$\thanks{E-mail: s.kaviraj@imperial.ac.uk}, Kok-Meng Tan$^{3}$, Richard S. Ellis$^{3,4}$ and Joseph Silk$^{3}$\\
$^{1}$Blackett Laboratory, Imperial College London, South Kensington Campus, London SW7 2AZ, UK\\
$^{2}$Mullard Space Science Laboratory, University College London, Holmbury St. Mary, Dorking, Surrey, RH5 6NT, UK\\
$^{3}$Department of Physics, University of Oxford, Keble Road, Oxford, OX1 3RH, UK\\
$^{4}$California Institute of Technology, 105-24 Astronomy,
Pasadena, CA 91125, USA\\}

\begin{document}

\pagerange{\pageref{firstpage}--\pageref{lastpage}} \pubyear{2008}

\maketitle

\label{firstpage}

%.........................................................................................................

\begin{abstract}
We exploit multi-wavelength photometry of early-type galaxies
(ETGs) in the COSMOS survey to demonstrate that the low-level star
formation activity in the ETG population {\color{black}at
intermediate redshift is likely to be driven by minor mergers.}
Splitting the ETGs into galaxies that show disturbed morphologies
indicative of recent merging and those that appear relaxed, we
find that $\sim32$\% of the ETG population appears to be
morphologically disturbed. While the relaxed objects are almost
entirely contained within the UV red sequence, their
morphologically disturbed counterparts dominate the scatter to
blue UV colours, regardless of luminosity.
{\color{black}Empirically and theoretically determined
major-merger rates} in the redshift range $z<1$ are several times
too low to account for the fraction of disturbed ETGs in our
sample, suggesting that minor mergers represent the principal
mechanism driving the observed star formation activity in our
sample. The young stellar components forming in these events have
ages between 0.03 and 0.3 Myrs and typically contribute $\leq10$\%
of the stellar mass of the remnant. Together with recent work
which demonstrates that the structural evolution of nearby ETGs is
consistent with one or more minor mergers, our results indicate
that the overall evolution of massive ETGs {\color{black} may be
heavily influenced} by minor merging at late epochs and highlights
the need to systematically study this process in future
observational surveys.
\end{abstract}

%.........................................................................................................


%.........................................................................................................

\section{Introduction}
The discovery of widespread recent star formation (RSF) in
early-type galaxies (ETGs) has brought several changes to our
traditional understanding of how this important class of
astronomical objects has formed and evolved over time. A
substantial literature on ETGs, based mainly on studies that use
optical data, has convincingly established that the bulk of the
stellar population in luminous early-types (particularly in
clusters) have formed at high redshift ($z>1$). Several pieces of
observational evidence point to this fact, including the small
intrinsic scatter in the early-type `Fundamental Plane'
\citep[e.g.][]{Jorg1996,Saglia1997,
Forbes1998,Peebles2002,Franx1993,Franx1995,VD1996}, red optical
colours
\citep[e.g.][]{BLE92,Bender1997,Ellis97,Stanford98,Gladders98,VD2000,Bernardi2003}
and chemical evidence for relatively short ($<1$ Gyr) star
formation timescales in these systems, deduced from the
over-abundance of $\alpha$-elements
\citep[e.g.][]{Thomas1999,Trager2000b}.

The star formation activity that has taken place in these galaxies
over the latter half of the Universe (since $z\sim1$) is therefore
appreciably weaker than the primordial bursts that built up the
bulk of their stellar populations. A drawback of optical data is
its relatively low sensitivity to small amounts of RSF. Young
stellar populations, which are dominated by hot, massive,
main-sequence stars, output a substantial fraction of their flux
in the ultraviolet (UV) spectral ranges (shortward of
$\sim2500$\AA). However, while the impact of low-level RSF on the
optical spectrum is reasonably weak, a small mass fraction (a few
percent) of young ($<1$ Gyr old) stars strongly affects the
rest-frame UV (Kaviraj 2008). Furthermore, the UV remains largely
unaffected by the age-metallicity degeneracy \citep{Worthey1994}
that typically plagues optical analyses, making it a useful
photometric indicator of RSF \citep{Kaviraj2007b}.

\begin{figure}
$\begin{array}{c}
\includegraphics[width=\columnwidth]{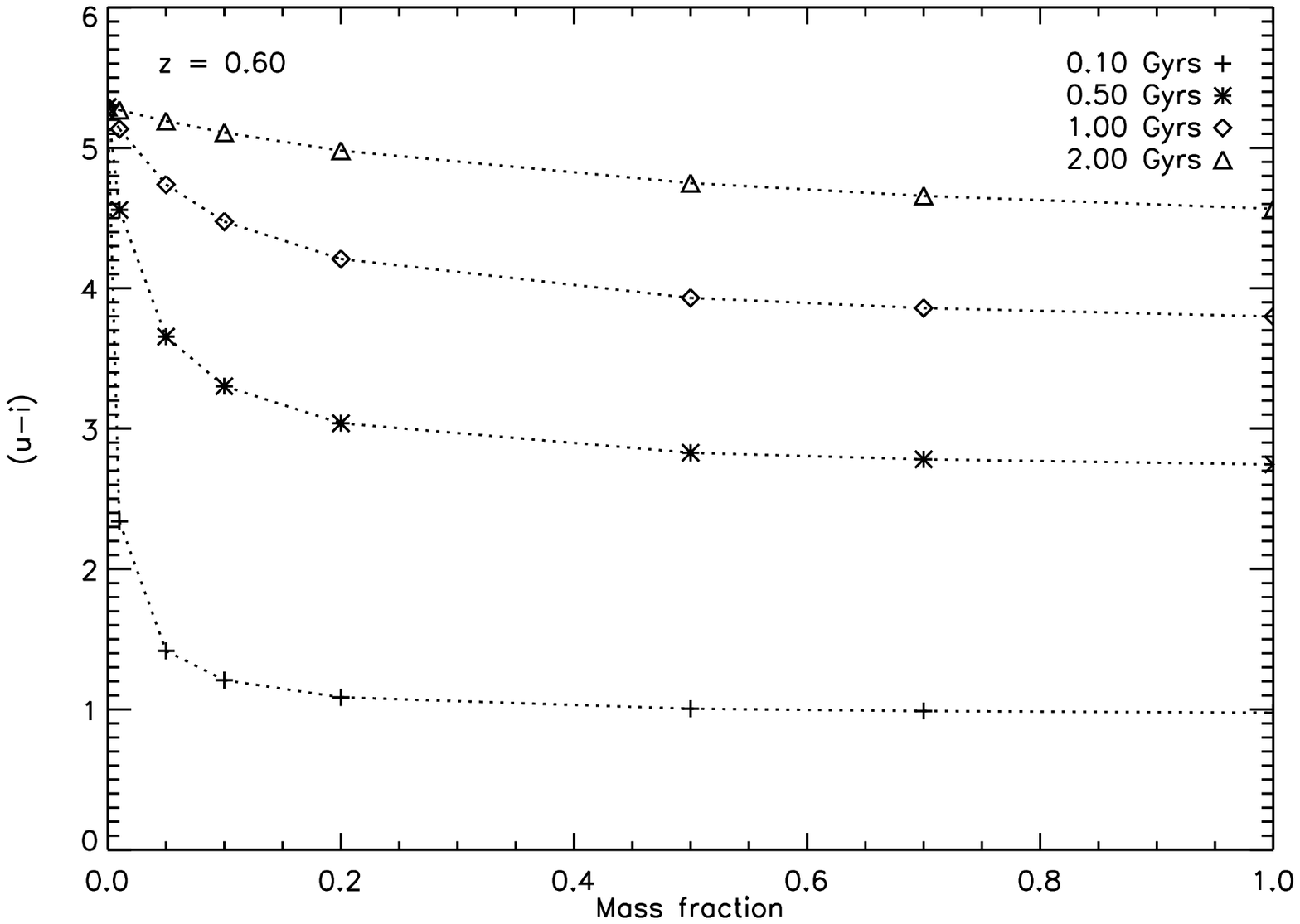}\\
\includegraphics[width=\columnwidth]{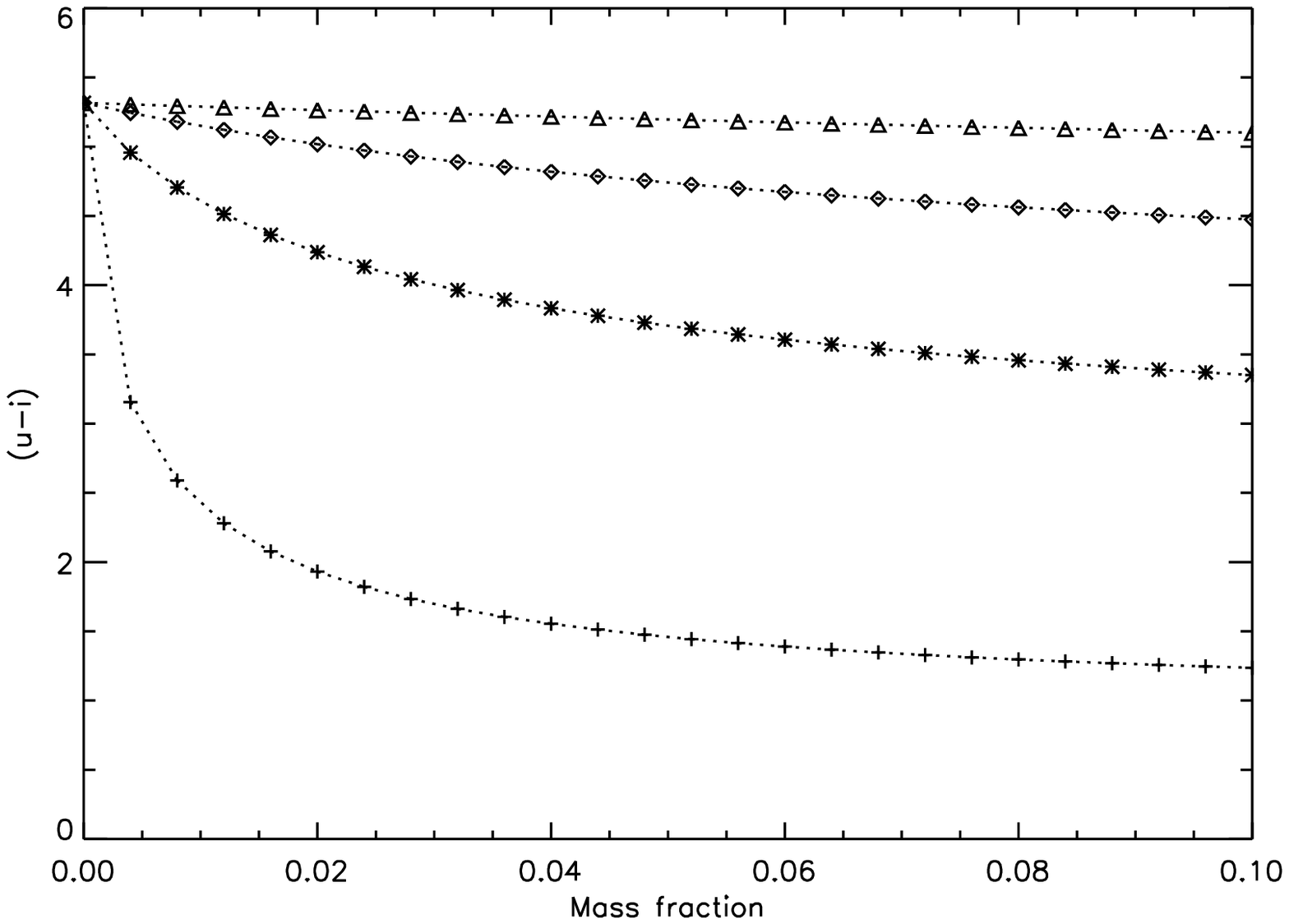}
\end{array}$
\caption{The sensitivity of the rest-frame UV to young stars. We
show the predicted $(u-i)$ colour (which traces rest-frame UV at
$z\sim0.6$) for small mass fractions of young stars superimposed
on an old underlying stellar population.{\color{black} The top
panel shows the full range of mass fraction values from 0 to 1,
while the bottom panel zooms in on mass fractions less than 0.1,
typical of minor mergers.} We assume two instantaneous bursts of
star formation, where the first burst is fixed at
{\color{black}$z=3$} and the second burst is allowed to vary in
age and mass fraction. The $(u-i)$ colour of the composite stellar
population, observed at $z\sim0.6$, is plotted as a function of
the age (symbol type) and mass fraction (x-axis) of the second
burst. It is apparent that even a small mass fraction ($\sim$1\%)
of young stars ($\sim0.1$ Gyrs old) causes a dramatic change in
the $(u-i)$ colour compared to what might be expected from a
purely old stellar population.} \label{fig:ui_tyfy}
\end{figure}

In Figure \ref{fig:ui_tyfy}, we demonstrate the sensitivity of the
rest-frame UV to various mass fractions of young stars. Since our
study will be based on intermediate-redshift ($z\sim 0.6$) data
drawn from the COSMOS survey (described in detail later in the
text), we present the analysis in the context of the COSMOS $u$
and $i$-band filters. For galaxies at $z \sim 0.6$, the observed
$(u-i)$ colour traces the rest-frame $(NUV-g)$ colour, where $NUV$
represents the near-UV spectrum ($\sim2500$\AA). In Figure
\ref{fig:ui_tyfy} we assume two instantaneous bursts of star
formation, where the first burst is fixed at $z=3$ and the second
burst is allowed to vary in age and mass fraction. The present-day
$(u-i)$ colour of the composite stellar population (emitted at
$z\sim0.6$) is plotted as a function of the age (symbol type) and
mass fraction (x-axis) of the second burst. The figure indicates
that even a small mass fraction ($\sim$1\%) of young stars
($\sim0.1$ Gyrs old) causes a significant change in the $(u-i)$
colour, compared to what might be expected from a purely old
stellar population. It is this sensitivity that allows a robust
identification of small, young stellar components that might be
mixed in with the old, underlying stellar populations that
dominate ETGs. Note that the UV flux fades relatively quickly
(within 2 Gyrs) because the massive UV-producing stars are also
short-lived. {\color{black}The stellar models used for this
analysis are described in Yi (2003)}.

The properties of the rest-frame UV have recently been exploited
in detail to study the presence of young stellar populations in
ETGs at low and intermediate redshifts, an expected consequence of
their evolution in the standard LCDM galaxy formation paradigm
\citep{Yi2005,Kaviraj2007b,Kaviraj2008b,Schawinski2007a}. These
studies have found that, in contrast to their optical colours,
luminous early-types show a large spread in their UV colour
distribution of almost 5 mags - a direct consequence of the
sensitivity of the UV to small amounts of RSF demonstrated in
Figure \ref{fig:ui_tyfy} above. The basic conclusion that can be
drawn from these efforts is that, while the bulk of the stellar
mass in the early-type population does indeed form at high
redshift ($z>1$), early-types of all luminosities form stars over
the lifetime of the Universe, with luminous systems
($-23<M_V<-21$) forming up to 10-15\% of their stellar mass after
$z=1$, with a tail to higher values. The persistent large scatter
in the rest-frame UV colours indicates widespread low-level star
formation in the early-type population over the last 8 billion
years (Kaviraj 2008).

Recent work that combines spatially resolved 2D analyses using
\emph{Hubble Space Telescope} (HST) images with spectroscopic data
are consistent with the findings from the rest-frame UV studies.
The high signal-to-noise (S/N) of HST imaging in fields such as
the Hubble Deep Fields and GOODS \citep{Giavalisco2004}, coupled
with the superb angular resolution of the images ($\sim0.05$
arcsec/pixel), have revealed that a substantial fraction ($>25$\%)
of morphologically selected ETGs exhibit blue optical cores
\citep{Menanteau2001a,Ferreras2005}. Not unexpectedly, such blue
cores are typically accompanied by emission and absorption lines
characteristic of recent star formation \citep{Ellis2001} and a
comparison to models suggests that the accretion rates are
typically $\sim10$\% by mass over 1 Gyr. The number fractions of
ETGs that carry signatures of blue cores and the mass fractions
estimated to be forming in the RSF events are remarkably similar
to the values found in studies that have used the rest-frame UV.

While unambiguous signatures of RSF have been found in the
early-type population, the \emph{source} of that star formation
remains uncertain. RSF requires (cold) gas and there are several
channels that could provide this gas supply e.g. internal mass
loss from stellar winds and supernova (SN) ejecta, condensation
from hot gas reservoirs and accretion of gas due to major or minor
mergers. The principal aim of this paper is to study the role of
mergers in driving the low-level recent star formation observed in
the early-type population, with a view to exploring the overall
contribution of this channel {\color{black}(and its components
i.e. major vs minor mergers)} compared to other plausible sources
of RSF in early-type galaxies.

{\color{black}Before we begin, we briefly review recent work that
has implications for the role of merging in the evolution of ETGs
at low and intermediate redshift. While theory and observation
both indicate that major mergers are central to the formation of
spheroids \citep[e.g.][]{Toomre_mergers,Bell2004,Faber2007},
recent work has highlighted the potentially influential role of
minor merging (where the mass ratios are $\leq$ 1:3) on ETG
evolution at late epochs. Early work by \citet{Schweizer1990} and
\citet{Schweizer1992} showed that fine structure in nearby
early-type galaxies, plausibly the result of recent minor
interactions, was correlated with spectro-photometric evidence for
RSF. More recently, \citet{Kaviraj2009} (see also
\citealt{Bezanson2009,Serra2010,Hopkins2010}) have shown that the
UV-optical properties of local ETGs can be reproduced purely by
minor mergers between massive spheroids and gas-rich satellites,
without recourse to the (much less frequent) major merger channel
at present-day. In this work, numerical simulations of minor
mergers are convolved with the expected frequency of such events
in the standard model, to show that the predicted scatter in the
UV and optical colours matches quantitatively with the
observations. Note that, while minor mergers have been shown to be
inefficient at triggering strong starbursts
\citep[e.g.][]{Cox2008}, the star formation induced in ETGs at
late-epochs is at a low level, with the new stars forming a few
percent or less of the mass of the remnant. Indeed, the
simulations of \citet{Kaviraj2009} indicate that minor mergers are
able to drive the requisite level of star formation in
low-redshift ETGs. It is worth noting that the results of the Cox
et al. study are consistent with those from \citet{Kaviraj2009},
even though the assumed morphologies of the merger progenitors in
the two sets of simulations are different. In particular, the mass
fraction of new stars for merger mass ratios of around 1:10 is
around 10\% (see Eqn. 5 in Cox et al.), which is consistent with
the RSF fractions reported in the recent literature
\citep[e.g.][]{Kaviraj2007b} and the mass fractions produced in
the simulations of \citet{Kaviraj2009}.

Using numerical simulations, \citet{Bournaud2007} and
\citet{Naab2009} have demonstrated that repeated minor merging can
lead to the formation of ETGs with morphological and kinematical
properties that are consistent with those observed in local ETGs.
The plausibility of these theoretical results is bolstered by the
fact that minor mergers are predicted and observed to be more
frequent than their major counterparts at late epochs \citep[see
e.g.][]{Stewart2008,Jogee2009,Lopez2010,Darg2010}. For example,
\citet[][see their Fig 6]{Stewart2008} show that $\sim$17\% of
haloes with mass $10^{12}$ M$_{\odot}$ have had a major merger
(mass ratio $>$ 1:3) since $z=1$, while the fraction of such
haloes that have experienced a minor merger is $\sim$55\%. Minor
mergers appear to be a factor of $\sim$3 more frequent, consistent
with similar observational estimates
\citep[e.g.][]{Jogee2009,Lopez2010}.

Indeed very deep optical imaging indicates that the fraction of
local ETGs that carry morphological disturbances
\citep[e.g.][]{VD2005,Tal2009} and associated RSF
\citep{Kaviraj2010b} is $\sim73$\%. Given typical major merger
timescales of $\sim0.4$ Gyrs
\citep[e.g.][]{Naab2006,Bell2006,Lotz2010}, it is reasonable to
suggest that the vast majority of \emph{local} ETGs have
experienced recent minor merging. The specific results on ETGs are
consistent with an emerging literature which has demonstrated the
diminishing impact of major merger activity on the evolution of
massive galaxies in general at late epochs ($z<1$) and a
corresponding increase in the significance of minor mergers. By
considering close pairs and merger remnants, \citet{Robaina2009}
studied the contribution of major mergers to the cosmic SFR at
intermediate redshift ($z<0.6$). They found that less than 10\% of
the star formation in massive galaxies at these redshifts are
plausibly triggered by major interactions. Similar trends have
been reported by \citet{Jogee2009} and \citet{Lopez2009} and, in a
narrower redshift range, by \citet{Wolf2005} and \citet{Bell2005}.

While local ETGs appear to show morphological disturbances
accompanied by spectro-photometric signatures of RSF, it is
important to establish and extend these results to higher
redshift. The advantages of studies at higher redshift are that
larger galaxy samples can be studied using the existing survey
data and morphological peculiarities can be detected more
efficiently using existing deep HST imaging at higher resolution
than is possible with ground-based telescopes at low redshift. For
example, the work of \citet{Kaviraj2009}, which provides a strong
plausibility argument for the role of minor mergers in ETG, has a
largely \emph{theoretical} basis, since standard Sloan Digital Sky
Survey \citep[SDSS; ][]{SDSSDR6} images are not deep enough and
lack the angular resolution to reveal morphological disturbances
\citep[see e.g.][]{Kaviraj2010a}. In a similar vein, while a
statistically significant correlation is found between disturbed
morphologies and RSF in \citet{Kaviraj2010b}, this study is based
on a small sample of $\sim$ 100 objects with very deep optical
imaging. There is a clear need to push these local results to
higher redshift, where the merger rates \citep{Khochfar2001} and
galaxy gas fractions \citep{Rao1994} are both expected to be
higher and one would expect a stronger and clearer result than the
studies in the local Universe.

The aim of this paper is, therefore, to probe the coincidence
between morphological disturbances and the presence of RSF in the
intermediate-redshift ETG population. To achieve these aims, we
require (a) photometric data which traces the rest-frame UV
spectral ranges (since the mass fractions of young stars are
expected to be small) and (b) deep, high-resolution imaging that
enables accurate visual morphological classification and
identification of features typically generated by mergers (e.g.
tidal tails and asymmetries). To this end we exploit publicly
available data from the multi-wavelength COSMOS survey to explore
the early-type population at intermediate redshifts ($0.5<z<0.7$).
A brief description of the COSMOS survey and the sample of
galaxies that underpins this study is given in the next section.
The main thrust of this work is two-fold. Firstly, we exploit the
COSMOS $u$-band filter to explore the \emph{rest-frame UV}
properties of ETGs in this field. Secondly, we specifically
separate ETGs that are morphologically relaxed from those that
carry signatures of recent merging and compare the rest-frame UV
properties of these two sub-populations.}

\begin{comment}
We note that several previous studies have explored the ETG
population at intermediate redshifts. While some of these works
have employed visual inspection of images
\citep[e.g.][]{Ferreras2005,Kaviraj2008}, others have used
parametric approaches such as the `CAS' \citep[see
e.g.][]{Conselice2003} system, $M_{20}$ or the Gini coefficient
(see e.g. Ferreras et al. 2005) to extract ETGs from large
observational surveys like COSMOS \citep{Scoville2007}. However,
these studies have focussed on the properties of the ETG
population as a whole and, unlike the work presented here, they
did not attempt to split the ETG population into galaxies that
appear relaxed and those that carry signatures of recent merging.
It is also worth noting that these morphological disturbances are
often subtle (see Section 2.2) and best identified through direct
visual inspection because parametric approaches are too coarse to
detect them.
\end{comment}

We note that a complication in the interpretation of excess UV
flux in \emph{low-redshift} ($z<0.2$) early-types is that the UV
contributors may be both young and old. {\color{black}Extreme
horizontal branch (EHB) stars and their progeny emit efficiently
in the UV spectrum and are thought to be largely responsible for
the `UV upturn' phenomenon in some massive early-type galaxies
{\color{black}\citep[e.g.][but see Han et al. (2003) for an
alternative source of UV upturn from sub-dwarf B and O
stars]{Yi97,Yi99,Yi2003}.} However, since the HB takes 8-9 Gyrs to
develop, the UV flux in galaxies at $z>0.5$ (such as those in this
study) will {\color{black}be negligibly affected} by the HB, since
the Universe would be too young for it to be in place.} Hence,
excess UV flux seen in early-type systems should be overwhelmingly
from young stars. By placing this study at intermediate redshift
we avoid having to disentangle the UV contributions of young and
old stellar populations, making the rest-frame UV flux a rather
clean photometric indicator of RSF.

This paper is structured as follows. In Section 2 we briefly
describe the COSMOS dataset used in this study and the process of
morphological classification. In Section 3 we compare the
morphologies of early-type galaxies and their rest-frame UV
properties. In Section 4 we use the multi-wavelength photometry
provided by COSMOS to quantify the properties of the RSF (ages,
mass fractions, metallicities and dust contents) in the early-type
population. In Section 5 we discuss the various plausible sources
of gas that could drive the residual star formation seen in ETGs,
discuss the role of merging in this context and speculate on what
the dominant RSF channel is likely to be in ETGs at late epochs.
Finally, in Section 6 we summarise the results of this study.
Cosmological parameters used for the background cosmology are
taken from the three-year WMAP observations \citep{Spergel2007}:
$\Omega_m = 0.241$, $\Omega_{\Lambda} = 0.759$, $h=0.732$,
$\sigma_8 = 0.761$.

\begin{figure*}
\begin{minipage}{172mm}
\begin{center}
\includegraphics[width=\textwidth]{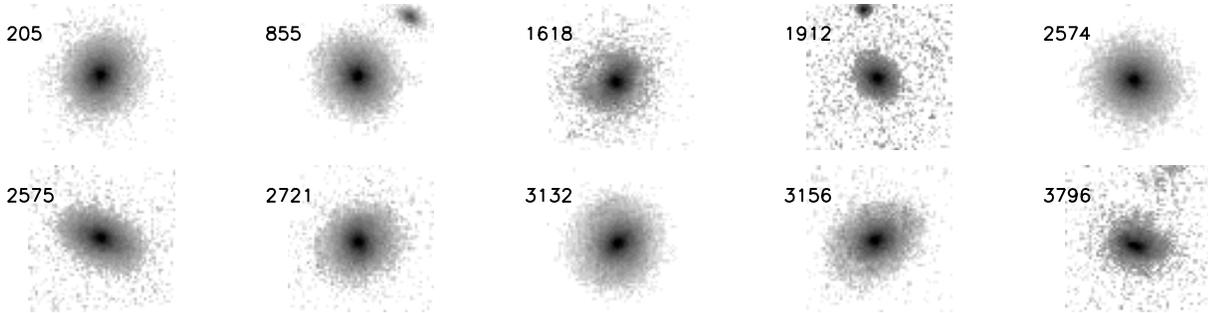}
\caption{Examples of galaxies that were classified as
\emph{relaxed ETGs} across the entire magnitude range considered
in this study ($i<22.3$). Objects in this category are typically
identified through their strongly peaked `de Vaucouleurs' light
profiles, coupled with smooth light distributions and a lack of
internal structure. Note that objects 1618, 3156 and 3796 are
examples of relaxed ETGs that reside in the UV blue cloud and are
discussed further in Section 5. All galaxy images are 3" on a
side.} \label{fig:etgrelaxed_examples}
\end{center}
\end{minipage}
\end{figure*}

\begin{figure*}
\begin{minipage}{172mm}
\begin{center}
\includegraphics[width=\textwidth]{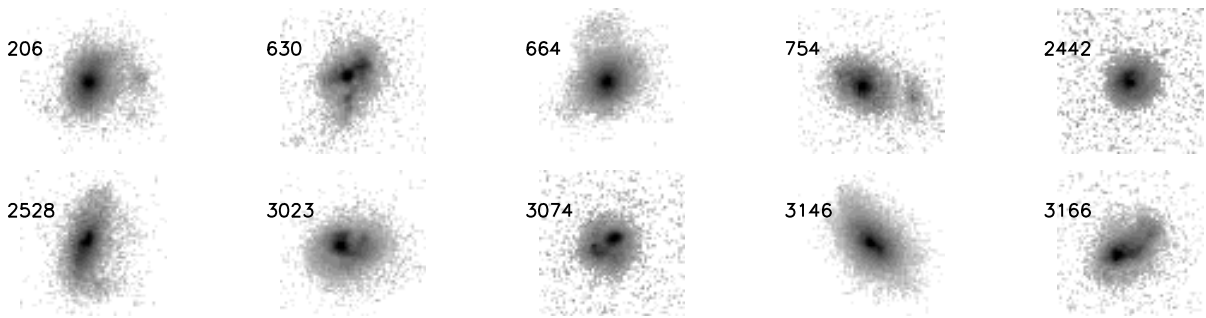}
\caption{Examples of galaxies that were classified as
\emph{disturbed ETGs} across the entire magnitude range considered
in this study ($i<22.3$). Note that all objects in this category
have bulge-like light distributions and smooth light distributions
in the main body of the galaxy. While some of the objects have
relaxed cores and exhibit disturbances only in the outer regions
of the galaxy (e.g. 206 and 664), others appear to still be in the
process of assimilating material from a recent merger (e.g. 630,
2442, 3023, 3074 and 3146). Tidal tails are visible in 2528 and
3166 and all the objects show asymmetries. All galaxy images are
3" on a side.} \label{fig:etgmerger_examples}
\end{center}
\end{minipage}
\end{figure*}

\begin{figure*}
\begin{minipage}{172mm}
\begin{center}
\includegraphics[width=\textwidth]{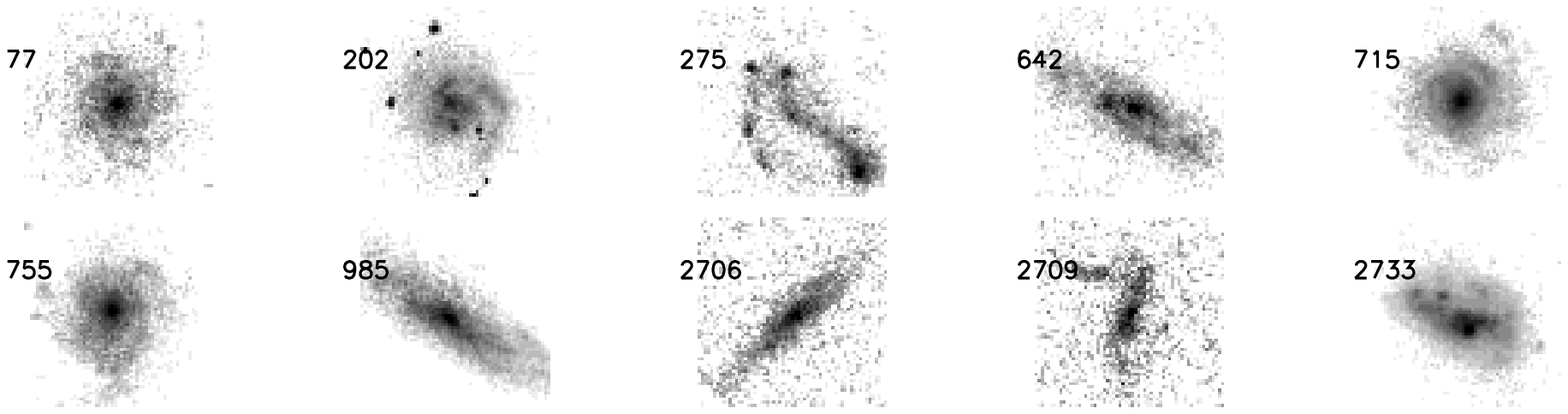}
\caption{Examples of galaxies classified as \emph{late-types
(LTGs)}. We purposely show several examples of face-on spirals
(77, 202, 715 and 755) that are most likely to be confused with
early-type galaxies in any method of morphological classification.
It is apparent that, within the magnitude range considered in this
study ($i<22.3$), the high angular resolution of the HST images
reveals internal structure in the face-on spirals, allowing them
to be differentiated from ETGs (in which the light distributions
are much smoother). This category includes relaxed spirals (such
as 77 and 715), irregulars (such as 275 and 2709) and disturbed
late-types (such as 642 and 2733). For the purposes of this study
these subdivisions are collated since we are only interested in
objects that are either relaxed early-types or will eventually
settle into an early-type morphology after signs of the merger or
interaction disappear. All galaxy images are 3" on a side.}
\label{fig:ltg_examples}
\end{center}
\end{minipage}
\end{figure*}

%.........................................................................................................

\section{Data}
The Cosmological Evolution Survey \citep[COSMOS;][]{Scoville2007}
is a deep, 2 deg$^2$ panchromatic survey offering imaging data and
photometry in 15 broadband filters between $0.3\mu$m ($u$-band)
and $2.4\mu$m ($K_s$-band), taken using the Subaru 8.3m telescope,
the KPNO and CTIO 4m telescopes, and the CFHT 3.6m telescope. The
entire COSMOS field (centred at 10:00:28.6, +02:12:21) has been
imaged by the HST Advanced Camera for Surveys (ACS) in the F814W
($i$-band) filter.

This study is based on a publicly available $i$-band selected
multi-wavelength catalog, based on first release optical and
near-infrared (NIR) data from COSMOS. In this paper, we combine
multi-band photometry \citep[][C07 hereafter]{Capak2007}, with
photometric redshifts \citep[][M07 hereafter]{Mobasher2007} and
HST images of COSMOS galaxies \citep{Koekemoer2007}.
{\color{black}To calculate the photometric redshifts and absolute
$V$-band magnitudes M07 have convolved a library of template
spectra, spanning the full range of local galaxy spectral types,
through the COSMOS filter curves. This library is then shifted in
redshift space and compared to the observed photometry of galaxies
by minimising the $\chi^2$ function (see M07 for more details).
The photometric redshifts have an accuracy of $dz/1+z\sim0.031$.
For the purposes of this study, the accuracy offered by the
photometric redshifts is sufficient and spectroscopic redshift
measurements are not required.} Finally, we note that the
recommended photometric offsets in C07 have been applied to each
filter and, where appropriate, total magnitudes have been
calculated by applying the offsets between the measured 3"
aperture magnitudes and total magnitudes that are provided in the
public catalogue.

%.........................................................................................................

\subsection{The sample}
Our study is based on a parent sample of 4086 galaxies in the
COSMOS field, brighter than a magnitude limit $i_{AB}<$22.3 with
redshifts in the range 0.5$<z<$0.7. {\color{black} The mean
redshift is $z=0.58$. These magnitude and redshift ranges imply
that our galaxies are brighter than $M_V<-20.5$ and have stellar
masses around $10^{10}$ M$_{\odot}$ or greater
\citep{Mobasher2007}.} The magnitude and redshift selection was
made to optimize the utility of rest-frame $NUV$ coverage (since,
at $z\sim$0.5, the optical $u$-band filter samples $\sim$2500\AA).
{\color{black}As we describe in Section 2.2 below, the magnitude
and redshift limits allow reliable classification of galaxy
morphologies through visual inspection of the HST F814W images.}
Finally, at these epochs, we expect limited contamination of the
UV flux by old HB stars, given that the maximum age of stellar
populations in our chosen redshift range is $\sim$ 5 to 6.5 Gyrs
(assuming star formation commenced around $z\sim$3).

%.........................................................................................................

\subsection{{\color{black}Morphological classification using visual
inspection}} {\color{black}Galaxy morphologies are classified in
this study using direct visual inspection of the F814W images,
which have a surface brightness limit of 27.2 mag arcsec$^{-2}$
\citep{Koekemoer2007}. At the redshifts probed by this study this
translates roughly into rest-frame $5100$\AA (approximately
rest-frame $V$-band). Several methods have been developed recently
that use automated parameters (e.g. concentration, asymmetry,
clumpiness, M$_{20}$ and the Gini coefficient) to separate the
galaxy population into broad morphological classes
\citep[e.g.][]{Abraham1996,Abraham2003,Conselice2003,Lotz2004}.
The performance of these methods are typically calibrated against
results from direct visual inspection, which offers better
precision and reliability in the classification of galaxy
morphologies \citep[e.g.][]{Lisker2008,Robaina2009}.

Several studies have exploited visual inspection to classify
galaxy morphologies, at similar redshifts, magnitude limits and
rest-frame wavelengths as those studied in this paper \citep[see
e.g.][]{Elmegreen2005,Ferreras2005,Bundy2005,Cassata2005,Jogee2009,Robaina2009}.
For example, \citet{Bundy2005} have performed visual inspection
using F850LP images for galaxies brighter than $m_z<22.5$ out to
$z \sim 1$ (in contrast to our sample which stops at $z=0.7$).
\citet{Robaina2009} have used images in the F606W filter to
classify galaxies in the redshift range $0.4<z<0.8$, while
\citet{Ferreras2005} have used F775W images to classify galaxies
with $i<23$ (almost a magnitude fainter than our limit) out to
$z=1$.

Several authors
\citep[e.g.][]{Dickinson2000,Elmegreen2005,Wolf2005,Bell2005,Jogee2009}
have noted that visually inspected morphologies from HST images
are stable (at intermediate redshifts) across a wide range of
imaging depths and wavelengths. For example, \citet{Wolf2005},
\citet{Bell2005} and \citet{Jogee2009} suggest that morphological
classification yields similar results in the redshift range
$0.4<z<0.8$, whether it is carried out using the F606W or F850LP
filters. Furthermore, the results of classification using the
relatively shallow imaging from the GEMS survey are very similar
to the ones achieved using the (1.5 mags) deeper images from the
GOODS survey. The images used for morphological classification in
this study have been taken using the F814W filter (which is very
close to F850LP) and which are 1 mag deeper than GEMS. Thus, based
on the findings of studies discussed above, that have performed
accurate visual inspection using both shallower and shorter
wavelength images compared to this work, the classification
results in this study should be robust.

Some further points are worth noting, given our specific interest
in ETGs and, in particular, in the subset of ETGs that harbour
morphological disturbances. Firstly, studies on low-redshift
samples indicate that using automated methods (e.g. concentration)
alone may result in some contamination of the ETG population by
bulge-dominated systems such as face-on spirals \citep[see
e.g.][]{Kaviraj2007b}. It is important that such contaminants are
avoided in this study, since interlopers with blue UV colours
would potentially skew our results. Visual inspection is the most
reliable way to identify such contaminants, making this a vital
part of our study. Secondly, in a related work,
\citet{Cassata2005} have used visual inspection of HST images to
split the galaxy population into ETGs and LTGs, sub-dividing the
two populations into their relaxed and disturbed counterparts.
They also show that automated parameters lack the sensitivity to
separate relaxed ETGs from their disturbed counterparts - this
distinction again requires visual inspection.

Finally, we briefly note the mass ratios of mergers that should be
identifiable using the F814W images used in this study.
\citet{Kawata2006} have shown that a `dry' merger with a mass
ratio of 0.03 (i.e. roughly 1:30) creates tidal debris that is
visible in imaging with a surface brightness limit of $\sim$27 mag
arcsec$^{-2}$ in the $R$-band. The surface brightness of the tidal
features appear to be around 24-26 mag arcsec$^{-2}$ (e.g. Figure
1 in this paper). Taking into account (1) the better throughput of
the F814W filter compared to the Cousins $R$-band filter ($\sim$
0.5 mags for a 9 Gyr population of solar metallicity) and (2) the
$(1+z)^4$ cosmological dimming between the mean redshift
($z\sim0.3$) of the simulated galaxy in Kawata et al. and the mean
redshift ($z\sim0.6$) of our sample ($\sim$1.2 mags), we find that
the F814W images are deep enough to detect such features (although
the detection of some of these features will be marginal). Given
that the strength of the tidal features typically increase with
the mass ratio of the merger \citep[e.g.][]{Feldmann2008}, mergers
with mass ratios greater than 1:30 should be visible in the COSMOS
images. Similarly, \citet{Feldmann2008} have explored a larger set
of mergers with various mass ratios (1:1 to 1:10) between (1)
massive ETGs and small spheroids (similar in spirit to Kawata et
al.) and (2) massive ETGs and small disks. Figure 3 in this study
indicates that the surface brightnesses of tidal features produced
by such mergers vary in the range 23-25 mag arcsec$^{-2}$.
Factoring in cosmological dimming ($\sim$1.8 mags) between the
redshift of their synthetic images ($z \sim 0.1$) and the mean
redshift of our sample ($z \sim 0.6$) and the difference in the
filter throughputs between Cousins $R$-band and F814W for an old
population, we conclude that merger remnants formed from
progenitors with mass ratios greater than 1:10 are very likely to
be visible in the COSMOS F814W images used in this study
\citep[see also][]{Jogee2009}.}\\

{\color{black}The visual inspection in this study splits the
galaxy
sample into three categories:}\\

\noindent \textbf{\emph{Relaxed ETGs}}\\
These are ETGs which appear to have no morphological disturbances
at the depth and resolution of the COSMOS HST images.
{\color{black}ETGs are identified from their smooth, centrally
concentrated light distributions and a lack of spiral structure.}
Figure \ref{fig:etgrelaxed_examples} shows examples of galaxies
classified as relaxed ETGs, across the entire magnitude range
considered in this study ($i<22.3$). All galaxy images shown in
this section are 3" on a side. {\color{black} Note that a minority
of the galaxies classed as relaxed ETGs do show hints of
asymmetry, although the disturbances are not large enough for us
to confidently classify these objects as disturbed. We show three
examples of such objects: 1618, 3156 and 3796. These objects
typically reside in the UV blue cloud, suggesting that they may
indeed be disturbed but that the tidal features may be below the
surface brightness limit of the images or are not easily visible
due to the orientation of the galaxy. We discuss this issue
further in Section 5.1.}\\

\noindent \textbf{\emph{Disturbed ETGs}}\\
These are ETGs which either exhibit morphological disturbances
indicative of a recent merger or are in the process of undergoing
such a merger. Strictly speaking these objects are, to high
confidence, ETG \emph{progenitors}, which will settle into
early-type morphology after signs of the merger disappear. Figure
\ref{fig:etgmerger_examples} shows typical examples of galaxies
classified as disturbed ETGs across the entire magnitude range
considered in this study ($i<22.3$). {\color{black}Note that,
similar to their relaxed counterparts, objects in this category
have smooth, centrally concentrated light distributions in the
main body of the galaxy.} While some of the objects have relaxed
cores and exhibit disturbances only in the outer regions of the
galaxy (e.g. 206 and 664), others appear to still be in the
process of assimilating material from a recent merger (e.g. 630,
2442, 3023 and 3074). Tidal tails are visible in 2528 and 3166.
While the main body of the disturbed ETGs resemble their relaxed
counterparts, they are all visibly asymmetric to different
degrees.\\

\noindent \textbf{\emph{Late-type galaxies (LTGs)}}\\
These are objects that do not fall within the above categories. We
present examples of such objects in Figure \ref{fig:ltg_examples}.
We purposely show several examples of face-on spirals (77, 202,
715 and 755) that are most likely to be confused with early-type
galaxies in any method of morphological classification. It is
apparent that, within the magnitude range considered in this study
($i<22.3$), the high angular resolution of the HST images reveals
the internal structure in the face-on spirals, allowing them to be
differentiated from ETGs (in which the light distributions are
much smoother). This category includes relaxed spirals (such as 77
and 715), irregulars (such as 275 and 2709) and disturbed
late-types (such as 642 and 2733). For the purposes of this study
these subdivisions are collated since we are only interested in
objects that are either relaxed early-types or will eventually
settle into an early-type morphology after signs of the merger
disappear.\\

The classification itself is performed in two stages. First, each
galaxy in the sample is visually inspected and assigned a
classification. Colour or magnitude information is not used during
this process to avoid biasing the results. This procedure is
repeated until the numbers in the disturbed ETG category begin to
converge i.e. the numbers in this category between classification
runs are within 1\% of each other. In our case, such convergence
was achieved after 4 such runs.

In the second stage, we perform a `comparative check' of the
images. We tile 25 random galaxies at a time, in which 10 are
relaxed ETGs or disturbed ETGs and the rest are LTGs. This step is
performed as a final check of the classifications and to correct
possible mistakes in marginal cases. It is potentially important
because the fine structure that we are looking for in the
disturbed ETGs is, in some cases, quite subtle and it is important
to ensure consistency throughout the classification process. For
example, in rare cases (3 objects) we found that an object
originally labelled as a disturbed ETG - when inspected next to a
set of relaxed ETGs - did not show enough morphological
disturbance to warrant being called disturbed. In this case the
classification was changed to a relaxed ETG. We did not find any
cases where the converse was true, neither was any re-labelling
required between the ETG and LTG categories.

The bulk of the classification effort was performed by SK, with
subsets verified by RSE. For the relaxed ETG and disturbed ETG
categories, the agreement between the classifiers was 99.1\%. The
fraction of galaxies brighter than $i\sim22.3$ that were
identified as either relaxed or disturbed ETGs was $\sim35\%$,
with the disturbed ETGs forming $\sim32\%$ of this population.
{\color{black}It is instructive to compare the results of our
visual inspection for the ETG population to previous studies. At
similar limits in redshift and magnitude, Bundy et al. (2005)
found their ETG fraction to be $\sim$33\%. Ferreras et al. (2005)
found a similar number of 34\%. Since the number density of
massive ETGs is seen to remain stable since $z \sim 1$
\citep{Ferreras2009}, it is useful to compare these values to
similar work in the low-redshift Universe. \citet{Schawinski2007b}
visually inspected 48,023 galaxies from a volume-limited sample
($M_r<-20.5$ and $z<0.1$) from the SDSS and found 15,729 ETGs,
giving an ETG fraction of $\sim$33\%. Similarly
\citet{Kaviraj2010a} used the deep $r$-band imaging in the SDSS
Stripe82 and found an ETG fraction of $\sim$ 34\% in the local
Universe ($z<0.05$). The Stripe82 images are $\sim$2 magnitudes
deeper than the standard SDSS imaging. Thus, the range of values
for the ETG fraction appears clustered around $\sim$33-34\% at
late epochs and our value of 35\% compares very well with previous
studies that have employed visual inspection, both in the local
Universe and at intermediate redshifts.

Finally, it is worth comparing the fraction of disturbed ETGs
($\sim$32\%) found in this study to other similar works in the
literature. \citet{Cassata2005} separate their ETG and LTG
populations into their relaxed and disturbed subclasses. Their
sample spans a much broader range of redshifts ($0<z<2$) than our
study and the sample size is more than a factor of 10 smaller (300
objects as opposed to $\sim$4000 galaxies in our sample).
Nevertheless, they find a disturbed ETG fraction of around 20\%
(the fraction of disturbed LTGs is 38\%). Assuming counting errors
this value could be between 16\% and 28\%. Noting that small tidal
disturbances will become harder to discern at higher redshifts
(because tidal features will be smaller and due to surface
brightness dimming), one would expect a smaller disturbed ETG
fraction in \citet{Cassata2005}. Coupled with the range of values
for the disturbed ETG fraction in this study (16-28\%), our
results seem consistent with this work. \citet{Jogee2009} suggest
that 68\% of massive galaxies (M $\geq$ 2.5 $\times$ 10$^{10}$
M$_{\odot}$) carry morphological disturbances indicative of a
recent merger in the redshift range $0.24<z<0.8$. Although their
estimate is higher, we note that the look-back time sampled by
Jogee et al. is 4 Gyrs as opposed to 1.5 Gyrs in this work, and
that, while the massive galaxy population may be dominated by
ETGs, Jogee et al. do not split their galaxy population by
morphology. Both these factors would be expected to result in a
higher value for the disturbed galaxy fraction. Finally,
\citet{Kaviraj2010a} found that the fraction of ETGs which carry
morphological peculiarities in the local Universe is around 25\%.
The lower surface brightness limit of the Stripe82 images
($\sim$26 mag arcsec$^{-2}$), coupled with the declining merger
rate \citep{Khochfar2006}, suggests that the lower disturbed ETG
fraction in \citet{Kaviraj2010a} is not inconsistent with the
findings of our study.}

%.........................................................................................................

\section{Morphology and rest-frame UV colours of the ETG population} 
\begin{figure*}
\begin{minipage}{172mm}
\begin{center}
$\begin{array}{c}
\includegraphics[width=0.8\textwidth]{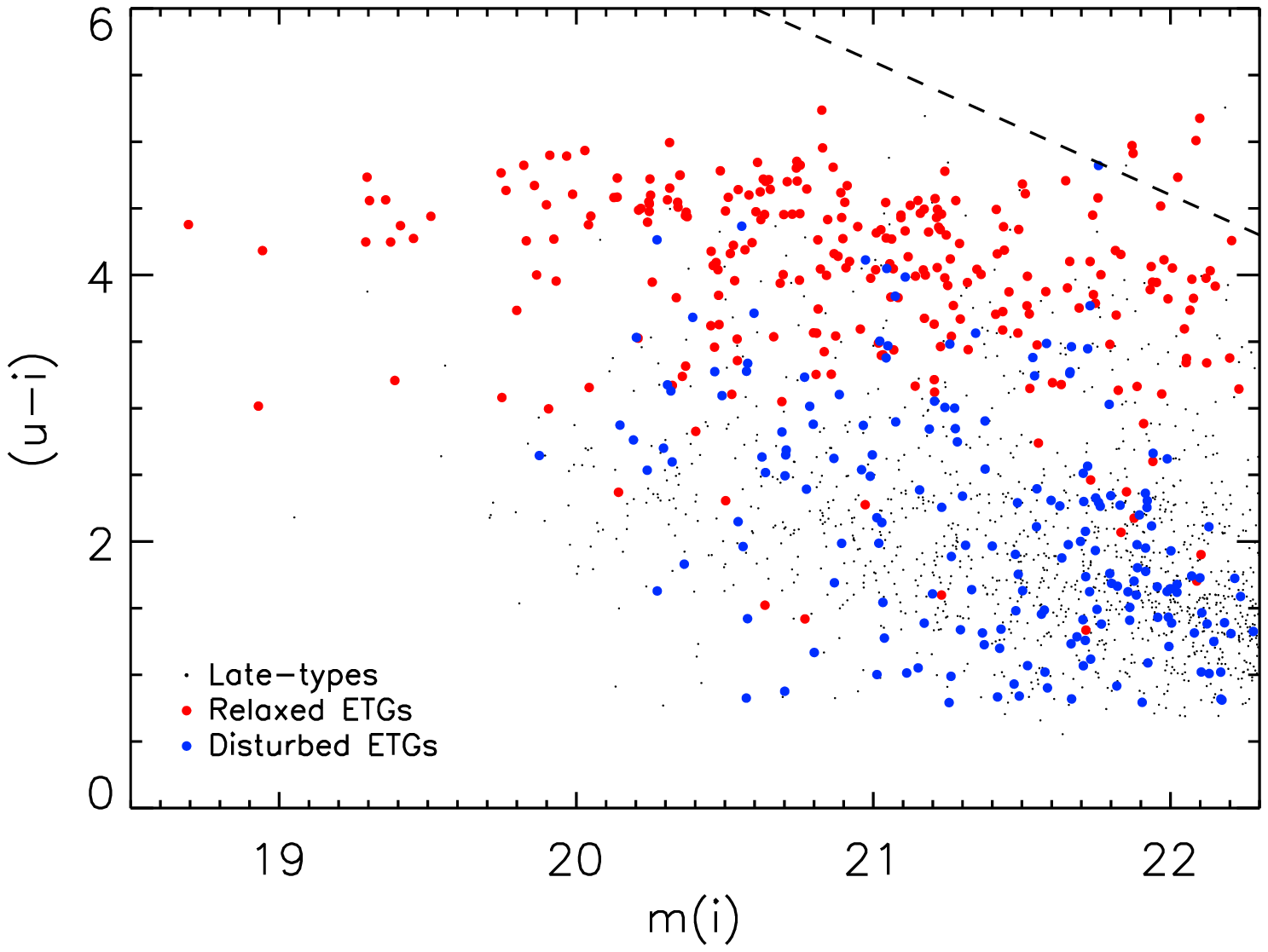}\\
\includegraphics[width=0.8\textwidth]{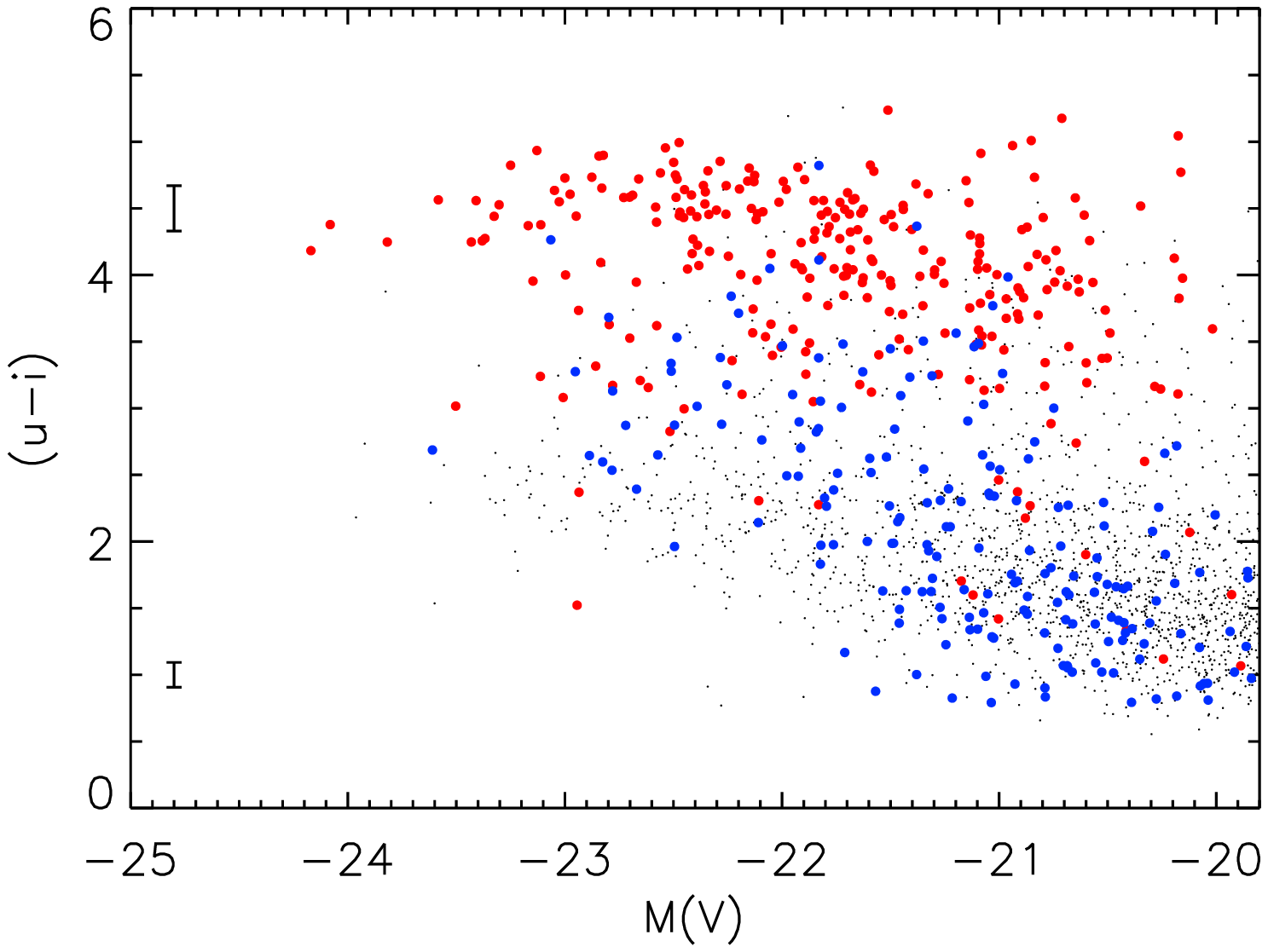}
\end{array}$
\caption{{\color{black}TOP: The $(u-i)$ colour-magnitude relation
of the early-type population. Filled red circles indicate galaxies
classified as relaxed ETGs, filled blue circles indicate disturbed
ETGs and black dots indicate the rest of the galaxy population.
The x-axis shows the $i$-band magnitude. The dashed line indicates
the detection limit based on the nominal depth of the $u$-band
filter ($m_{AB}=26.4$). BOTTOM: Same as the top panel but with the
absolute $V$-band magnitude of the galaxies plotted on the x-axis.
Note that, in this figure, absolute magnitudes were taken from the
catalogue of M07.}} \label{fig:colour_properties}
\end{center}
\end{minipage}
\end{figure*}

We begin our analysis, in Figure
\ref{fig:colour_properties}, by presenting the $(u-i)$
colour-magnitude relations (CMRs) of our galaxy sample, split by
their morphological types. Recall that the observed $(u-i)$
colours traces the rest-frame $(NUV-g)$ colour at these redshifts.
Red circles indicate relaxed ETGs, blue circles represent
disturbed ETGs and black dots indicate the rest of the galaxy
population. Galaxies separate into a broad UV red sequence and
blue cloud around $(u-i)\sim3$.

In the top panel of Figure \ref{fig:colour_properties} we show the
$(u-i)$ colours as a function of the observed $i$-band magnitudes.
The dashed line indicates the detection limit, based on the
nominal $u$-band filter depth ($m_{AB}=26.4$). The UV red sequence
is detectable across almost the entire magnitude range of interest
in this paper ($i<22.3$). In the bottom panel of this figure we
show the $(u-i)$ CMR as a function of the absolute $V$-band
magnitude of the galaxies.

\begin{comment}
The threshold below which the visual classification process did
not label any objects as ETGs ($i\sim 22.3$) is indicated by the
dotted line. It should be noted that the lack of ETGs at
magnitudes fainter than $i\sim22.3$ is simply an artifact of the
classification process and not an inherent property of the galaxy
population. While ETGs clearly do exist in the magnitude range
$i>22.3$, the decreasing S/N of the images prevents us from
identifying them with confidence because the smoothness of the
light distributions cannot be guaranteed in the noisier images.
\end{comment}

\begin{figure*}
\begin{minipage}{172mm}
\begin{center}
$\begin{array}{c}
\includegraphics[width=0.8\textwidth]{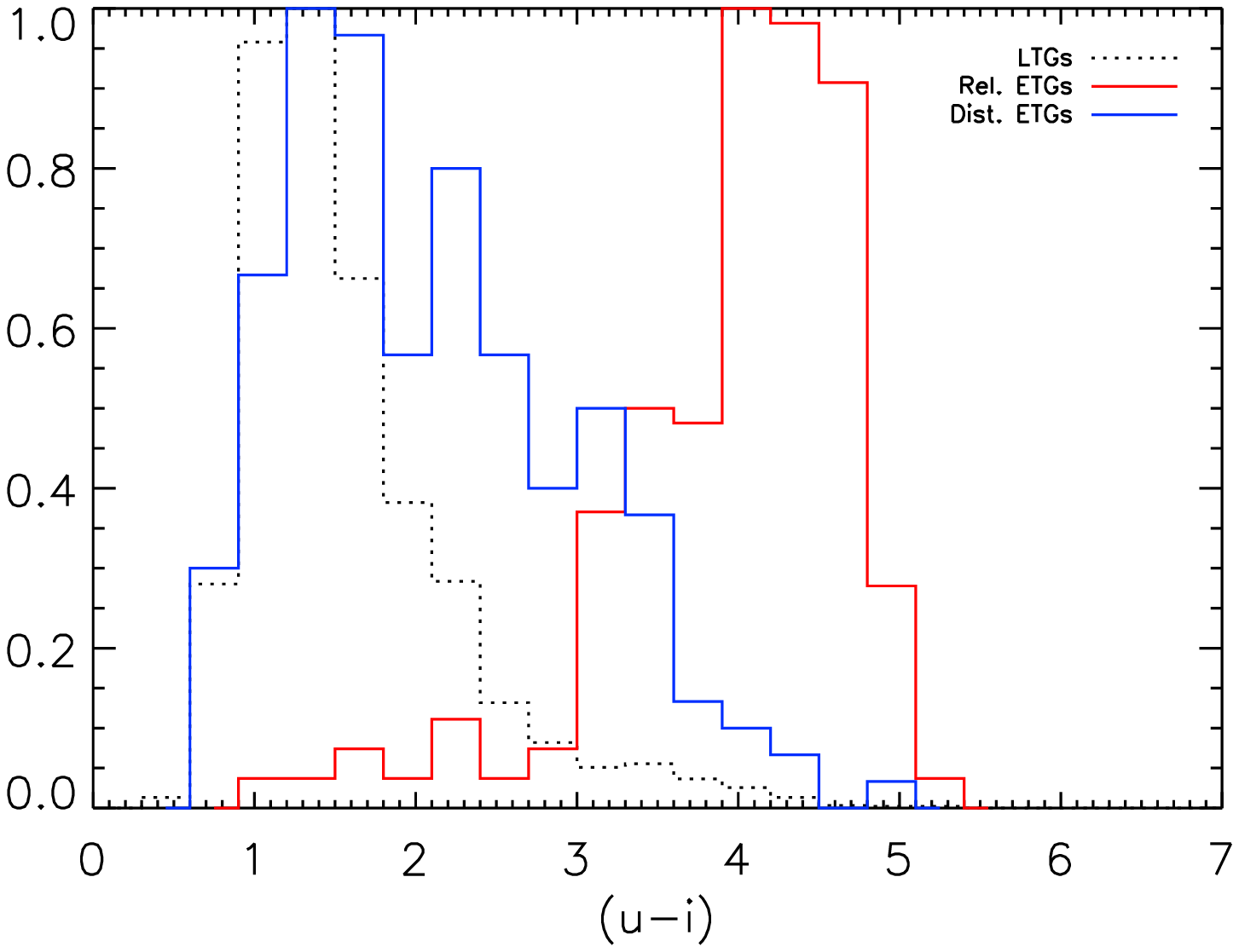}
\end{array}$
\caption{$(u-i)$ colour histograms for the relaxed ETG, disturbed
ETG and late-type populations. Note that all histograms are
normalised to 1.} \label{fig:colour_histograms}
\end{center}
\end{minipage}
\end{figure*}

The $(u-i)$ CMR indicates that the relaxed and disturbed ETGs are
reasonably well-separated in the rest-frame UV colour. The
separation is shown more explicitly in the top panel of Figure
\ref{fig:colour_histograms} which shows that, while the relaxed
ETGs peak in the UV red sequence with a minor tail in the UV blue
cloud ($u-i<3$), the disturbed ETG population peaks in the UV blue
cloud with a significant tail into the UV red sequence. Not
unexpectedly, the late-type population is almost completely
contained within the UV blue cloud. {\color{black}Note that this
pattern persists if the galaxy population is split into luminosity
bins.}

\begin{figure}
$\begin{array}{c}
\includegraphics[width=\columnwidth]{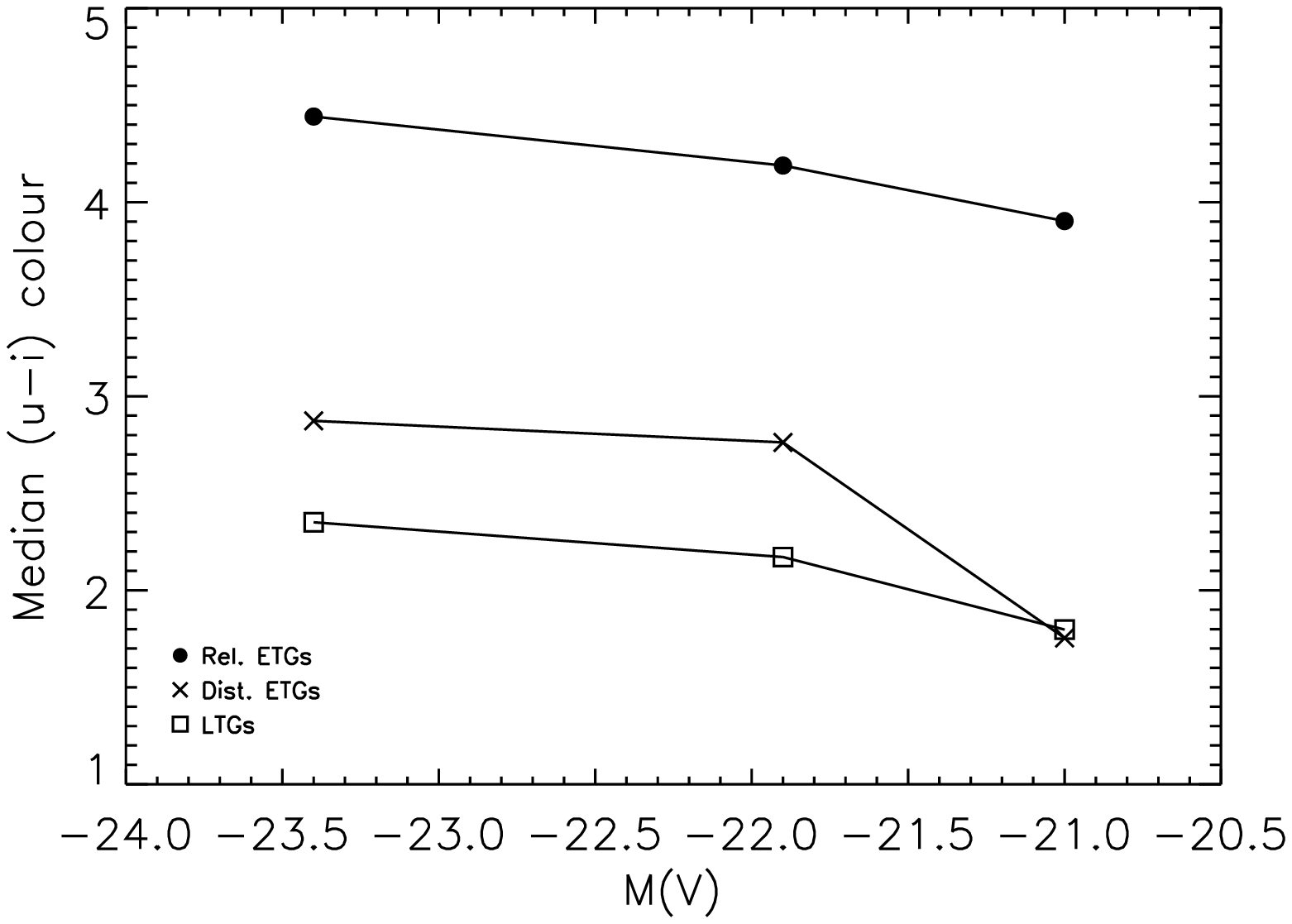}\\
\includegraphics[width=\columnwidth]{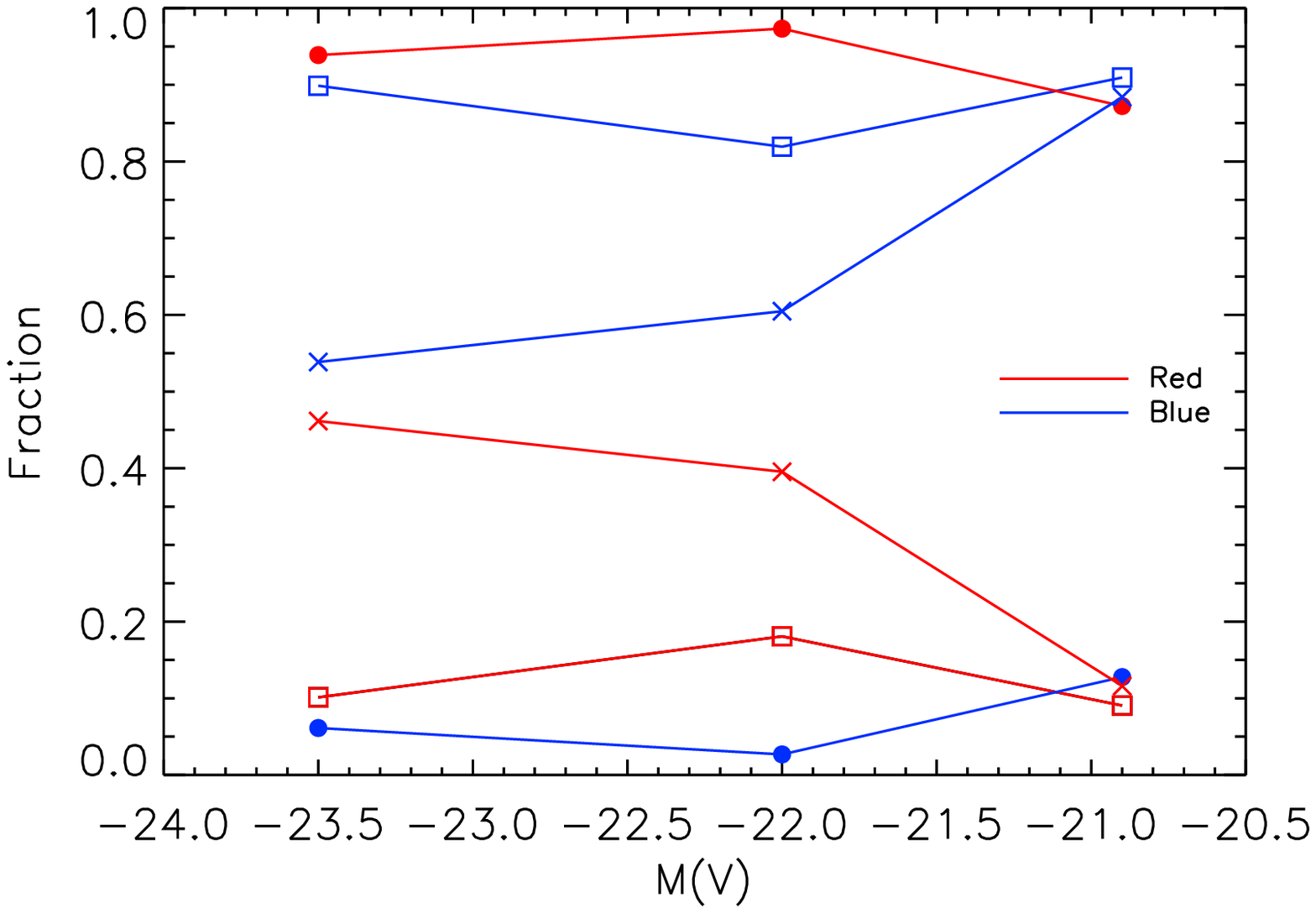}
\end{array}$
\caption{{\color{black}TOP: Median $(u-i)$ colours of relaxed ETGs
(filled circles), disturbed ETGs (crosses) and late-types (open
squares) in the three luminosity bins used in Figure
\ref{fig:colour_histograms}. Note that the points are plotted at
the mid-points of each luminosity bin. BOTTOM: The fraction of
each population that lies on the UV red sequence (red lines) and
in UV blue cloud (blue lines).}} \label{fig:red_blue_fractions}
\end{figure}

The information in these colour distributions is summarised
further in Figure \ref{fig:red_blue_fractions}. In the top panel
we show the median $(u-i)$ colour of each morphological class in
three luminosity bins. Note that the points are plotted at the
mid-points of each luminosity bin. It is apparent that, regardless
of luminosity, ETGs that carry morphological signatures of merging
are at least 1.5 mags bluer than their relaxed counterparts in the
rest-frame UV colour. Luminous ($M_V<-21.5$) disturbed ETGs appear
redder in the $(u-i)$ colour than late-types of comparable
luminosity by $\sim 0.3$ mags, with the colours becoming
comparable in the lowest luminosity bin ($-21.5<M_V<-20.5$).

The bottom panel of this figure shows the fraction of each
morphological class that lies on the UV red sequence (red lines)
and the fraction that inhabits the UV blue cloud (blue lines). In
the two highest luminosity bins ($M_V<-21.5$) the relaxed ETGs
(filled circles) lie almost exclusively on the UV red sequence. In
the lowest luminosity bin the red fraction is lower (90\%) but the
overwhelming majority of relaxed ETGs still inhabit the red
sequence.

In contrast, the disturbed ETG population resides mainly in the UV
blue cloud, regardless of luminosity. In the two highest
luminosity bins, 60\% of the disturbed ETGs reside in the blue
cloud with this fraction climbing to 86\% in the lowest luminosity
bin. The abrupt change in behaviour in the lowest luminosity bin
($-21.5<M_V<-20.5$), both in terms of the median $(u-i)$ colour
and the red/blue fractions, is most probably driven by
incompleteness at the faint end of the red sequence. The top
panel of Figure \ref{fig:colour_properties} indicates that the
detection limit begins to encroach upon the red sequence
population around $i\sim22$, although it is worth noting that
detections at fainter magnitudes do exist. However, it is possible
that fewer red sequence objects are detected in the lowest
luminosity bin, reducing the red fractions in all morphological
types.

%.........................................................................................................

\section{Quantifying the star formation histories}
We proceed by quantifying the RSF history of the ETG population to
explore the ages and mass fractions of the young stellar
populations that are forming in the disturbed ETGs and study the
differences between the relaxed ETGs and their disturbed
counterparts. {\color{black}We estimate parameters governing the
star formation history (SFH) of each galaxy by comparing its
multi-wavelength COSMOS photometry to a library of synthetic
photometry, generated using a large collection of model SFHs that
are specifically designed for studying ETGs. The chosen
parametrisation describes the broad characteristics of ETG SFHs
with a minimum of free parameters. A key feature of the scheme is
that the RSF episode is decoupled from the star formation that
creates the bulk, underlying population.

Since the existing literature on ETGs demonstrates that the bulk
of the stellar mass in these galaxies is metal-rich and forms at
high redshift over short timescales (see introduction), we model
the underlying stellar population using an instantaneous burst at
high redshift. We put this first burst at $z=3$ and assume that it
has solar metallicity. The RSF episode is modelled by a second
instantaneous burst, which is allowed to vary in (a) age between
0.01 Gyrs and the look-back time corresponding to $z=3$ in the
rest-frame of the galaxy (b) mass fraction between 0 and 1 and (c)
metallicity between 0.05Z$_{\odot}$ and 2.5Z$_{\odot}$. In
addition, we allow a range of dust values, parametrised by
$E_{B-V}$, in the range 0 to 0.5. The dust is applied to the model
galaxy as a whole and the empirical law of \citet{Calzetti2000} is
adopted to calculate the dust-extincted SEDs. The free parameters
are the age ($t_2$), mass fraction ($f_2$) and metallicity ($Z_2$)
of the second burst and the dust content ($E_{B-V}$) of the
galaxy.

Note that putting the first burst at $z=2$, or even $z=1$, does
not affect our conclusions about the RSF, because the first burst
does not contribute to the UV, which is dominated by hot, massive
main sequence stars with short lifetimes. The UV decays after
around a Gyr (and has almost completely disappeared after $\sim$2
Gyrs) as the UV-producing stars come to the end to their
lifetimes. Recall that the highest redshift being sampled in this
study is $z \sim 0.7$ which corresponds to an age of $\sim$5 Gyrs
if star formation begins at $z=3$ in a standard cosmology. It is
worth noting that our parametrisation is similar to previous ones
used to study elliptical galaxies at low redshifts using
UV/optical photometry \citep[e.g.][]{Ferreras2000} and
spectroscopic line indices \citep[e.g.][]{Trager2000b}.}

To build a model library of synthetic photometry, each combination
of the free parameters is combined with the stellar models of Yi
(2003) and convolved with the correct COSMOS
($u,g,r,i,z,K_s$)\footnote{Note that the $u$-band filter is from
the CFHT Mega-Prime instrument, the $g,r,i,z$ filters are from the
Subaru Suprime-Cam instrument and the $K_s$ filter is from the
KPNO FLAMINGOS instrument.} filtercurves. The library contains
$\sim600,000$ individual models. Since our galaxy sample spans a
range in redshift, equivalent libraries are constructed at
redshift intervals $\delta z=0.02$ in the redshift range
$0.5<z<0.7$. Note that the stellar models assume a
\citet{Salpeter1955} initial mass function.

For each ETG in our sample, parameters are estimated by choosing
the model library that is closest to it in redshift and comparing
its ($u,g,r,i,z,K_s$) photometry to every model in the synthetic
library. The likelihood of each model ($\exp -\chi^2/2$) is
calculated using the value of $\chi^2$, computed in the standard
way. The error in $\chi^2$ is computed by adding, in quadrature,
the observational uncertainties in the COSMOS filters with errors
adopted for the stellar models, which we assume to be 0.05 mags
for each optical filter and 0.1 mags for the $K_s$ filter (see Yi
2003). From the joint probability distribution, each free
parameter is marginalised to extract its one-dimensional
probability distribution function (PDF). We take the median value
of the PDF to be the best estimate of the parameter in question.
The 25th and 75th quartile values provide an estimate of the
uncertainties in these estimates. This procedure yields, for every
galaxy in our sample, a best estimate and error for each free
parameter. Note that the accuracy of the photometric redshifts
provided in the catalogue is sufficient for accurate parameter
estimation. Past experience suggests that, given the degeneracies
within the parameter space, the added accuracy of spectroscopic
redshifts does not change the derived distributions of parameter
values in such a study.

{\color{black}It is worth noting that the presence of (Type II)
AGN will not affect our analysis of the UV colours or the derived
values of the RSF parameters (our sample does not contain
quasars). The contamination from a Type II AGN is likely to be
less than around 15\% in UV flux \citep{Salim2007}, which
translates to around 0.15 mags in the ($NUV-r$) colour, much
smaller than the observed spread (around 4 mags) in the UV
colour-magnitude relation (see Figure 5 above). Previous work on
ETGs using UV/optical data at low redshift indicates that blue ETG
colours are not restricted to galaxies hosting emission line AGN
\citep{Schawinski2007b}. The same analyses also show that the
quality of the SED fitting is equally good in galaxies which carry
emission-line signatures of AGN and those that do not show any
signs of AGN, indicating that there is no measurable contribution
from a power-law component in the UV and optical spectrum.
Finally, a study of the GALEX \citep{Martin2005} UV images of
nearby AGN hosts indicates that the UV emission is extended,
making it unlikely that it comes from a central source
\citep{Kauffmann2007}. In summary, the parameter estimation
performed in this study is immune to the presence of a Type II
AGN.}

\begin{figure}
$\begin{array}{c}
\includegraphics[width=\columnwidth]{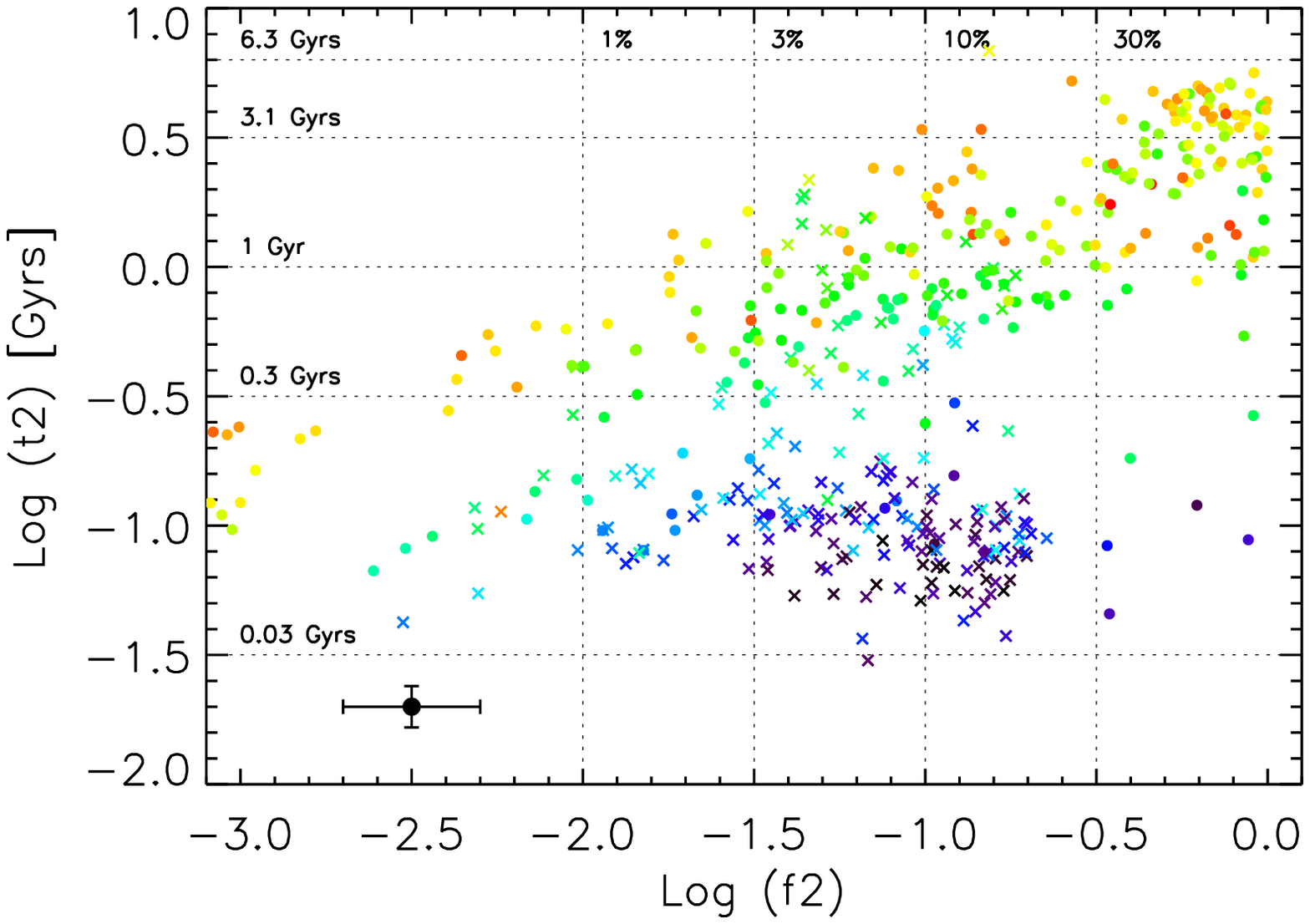}\\
\includegraphics[width=\columnwidth]{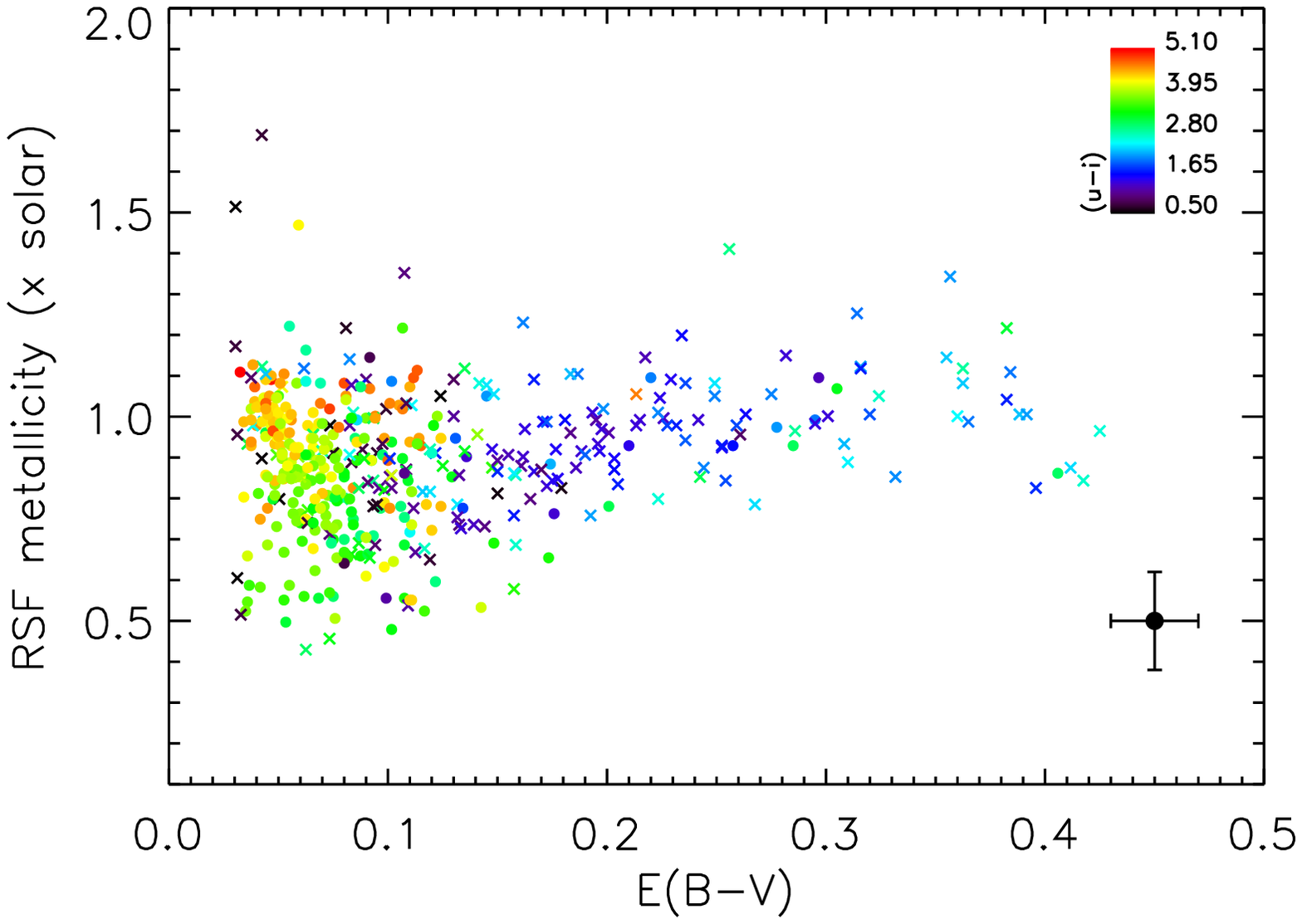}
\end{array}$
\caption{TOP: The $t_2-f_2$ space for the ETG population studied
in this paper. Recall that $t_2$ is an estimate of the age of the
recent star formation (RSF) and $f_2$ is its mass fraction.
BOTTOM: The metallicity of the RSF (y-axis) plotted against the
dust extinction in the galaxy (x-axis). Relaxed ETGs are shown
using filled circles and disturbed ETGs are shown using crosses.
Galaxies are colour-coded according to their $(u-i)$ colours.}
\label{fig:twoburst_parameters}
\end{figure}

In the top panel of Figure \ref{fig:twoburst_parameters} we
present the $t_2-f_2$ space for the ETG population studied in this
paper. Relaxed ETGs are shown using filled circles and disturbed
ETGs are shown using crosses. Galaxies are colour-coded according
to their $(u-i)$ colours. We find that the RSF ages in the
disturbed ETGs are typically between $\sim0.03$ Gyrs and $\sim0.3$
Gyrs. Not unexpectedly, the relaxed ETG population, which
predominantly resides on the UV red sequence, tends to occupy high
values of $t_2$. However, it is worth noting that, within errors,
the values of $t_2$ in the top panel of Figure
\ref{fig:twoburst_parameters} indicate that some relaxed ETGs,
especially those that are bluer than $(u-i)\sim4$, are likely to
contain intermediate-age (1-3 Gyr old) stellar populations. In
other words, while their photometry does not indicate the presence
of RSF, it is also inconsistent with their entire stellar
populations forming at high redshift. Such objects are therefore
likely to host intermediate-age stellar components.

The mass fractions formed in the RSF events are typically less
than 10\% (with a small tail to higher values). Note that the
median estimate for the mass fraction typically has a large error
(more than half a decade). The reason for these large
uncertainties becomes apparent if we refer back to Figure
\ref{fig:ui_tyfy}, which shows the evolution of the $(u-i)$ colour
as a function of the RSF mass fraction at a given RSF age. For all
RSF ages, increasing the mass fraction beyond a threshold value of
$\sim$10-20\% does not produce any further significant changes in
the $(u-i)$ colour. This is because, beyond a point, the spectral
energy distribution (SED) becomes dominated by the young stellar
component, so that increasing its mass fraction further only
changes the normalisation of the SED but not its shape (which
determines the colours). Hence, the same $(u-i)$ colour can be
consistent with a large range of mass fractions, producing a large
degeneracy in the mass fraction values. Note, however, that the
$(u-i)$ colour evolves rapidly with age regardless of mass
fraction, making the rest-frame UV a more robust indicator of the
RSF age ($t_2$) than the RSF mass fraction ($f_2$).

\begin{comment}
\begin{figure}
\includegraphics[width=\columnwidth]{rsfmet_Mv} \caption{TOP: RSF
metallicity plotted as a function of the absolute $V$-band
magnitude of the disturbed ETGs.} \label{fig:rsfmet_Mv}
\end{figure}
\end{comment}

In the bottom panel of this figure we present the remaining free
parameters in the analysis: the metallicity of the RSF and the
dust extinction applied to the galaxy. Similar to red early-types
at low redshift, the relaxed ETG population is typically
dust-poor, with $E_{B-V}$ values less than 0.1. Not unexpectedly,
the bulk of the disturbed ETGs are dustier ($0.1<E_{B-V}<0.4$),
since their star-forming regions (which dominate the UV fluxes)
are gas-rich and therefore expected to also contain dust. An
interesting result is that the RSF metallicities are typically
sub-solar but reasonably high, suggesting that the gas that forms
the young stars could already be metal-enriched. Recall that the
metallicity grid spans a wide range of values
(0.05-2.5Z$_{\odot}$) and very low metallicities are included in
the grid.

\begin{comment}
In particular, since the RSF is driven by the gas reservoirs in
the progenitors, this indicates that the \emph{gas-phase}
metallicities of the progenitors maybe reasonably high. Although
gas-phase metallicities have not been studied at intermediate
redshift, \citet{Tremonti2004} have performed such a study in the
local Universe. Table 2 in this study indicates that half-solar
gas-phase metallicities correspond to galaxies brighter than
$M_z\sim-19.5$, which roughly translates into $M_V\sim-18.5$. This
suggests that the progenitors involved that cause the disturbed
ETGs studied in this paper are typically brighter than this value.
\end{comment}

%.........................................................................................................

\section{What dominates the star formation
in the ETG population at late epochs?} The preponderance of
disturbed ETGs in the UV blue cloud indicates that merger-induced
star formation plays an important role in driving the residual
star formation activity observed in the ETG population. Luminous
ETGs, such as the ones studied here, could either have undergone
equal mass mergers (major mergers) or experienced satellite
accretion where the mass ratios are typically smaller than 1:3
(generically referred to as minor mergers).

Empirically determined major merger rates \citep[e.g.][Jogee et
al.
2009]{Patton2000,Lefevre2000,Patton2002,Conselice2003,Lin2004,Bell2006,Lotz2006,Conselice2007,Conselice2009,Bundy2009}
indicate that major merger activity in galaxies with masses such
as the ones considered in this paper becomes infrequent after
$z\sim1$. It is worth noting that while all of these works have
carefully explored the frequency of equal-mass mergers, some
discrepancy remains in the reported values for the major merger
rate, possibly driven by different selection techniques and cosmic
variance. {\color{black} While some studies
\citep[e.g.][]{Bell2006,Lotz2006,Conselice2007} indicate that each
massive galaxy undergoes $\sim$0.5 major mergers after $z\sim1$,
i.e. only 50\% of the massive galaxy population undergoes a major
merger since $z\sim1$, other efforts
\citep[e.g.][]{Lin2004,Lopez2009} suggest a lower value
($\sim9$\%). Similarly, \citet{Jogee2009} find that $\sim$16\% of
massive galaxies in the redshift range $0.4<z<0.8$ have undergone
a major merger. Given the uncertainty in these estimates it is
important to compare them to those derived from cosmological
simulations. The study of \citet{Stewart2008} indicates that the
fraction of massive haloes (with masses between 10$^{12}$ and
10$^{13}$ M$_{\odot}$) that have undergone a major merger (mass
ratios $>$ 1:3) since $z=1$ is around 20\% (left-hand panel of
Fig. 5 or right-hand panel of Fig. 6 in this paper). The
observations bracket this value, with the theoretical estimates
being closer to the \citet{Lin2004} and \citet{Lopez2009}
studies.} The fraction of ETGs that could be remnants of major
mergers in the redshift range probed by this study ($0.5<z<0.7$)
can be estimated as:

\begin{equation}
\frac{(\Delta T_{0.5-0.7})+0.4 \textnormal{(Gyrs)}}{7.71
\textnormal{(Gyrs)}}\times f_m,
\end{equation}

where $\Delta T_{0.5-0.7}$ ($\simeq1.42$ Gyrs) is the time elapsed
between $z=0.5$ and $z=0.7$, 7.71 Gyrs is the look-back time to
$z=1$, `0.4 Gyrs' is the visibility timescale of a major merger
\citep[e.g.][]{Naab2006,Bell2006,Lotz2010} and $f_m$ is the
fraction of galaxies that have experienced a major merger since
$z=1$. We find that adopting $f_m \sim 0.5$
\citep[e.g.][]{Bell2006,Lotz2006,Conselice2007,Conselice2009}
implies a major merger contribution of $\sim12$\% to the
{\color{black}disturbed} ETG population studied in this paper,
while adopting $f_m \sim 0.09$ from \citet{Lin2004} yields a
negligible value of $\sim2$\%. It is apparent that even the higher
value for the major merger contribution ($\sim12$\%) is not
sufficient to satisfy the observed disturbed ETG fraction
($\sim32$\%) found in our galaxy sample. {\color{black}Recall from
above that the theoretical estimates of the major merger rate
\citep[e.g.][]{Stewart2008} are lower than $f_m \sim 0.5$, so that
the major merger contribution could well be lower than
$\sim12$\%.}

{\color{black}Since the mass fractions formed in these events are
typically small, it is likely that any major mergers that do take
place do not lead to significant star formation. This suggests, in
agreement with recent theoretical work \citep{Khochfar2009}, that
the progenitors of major mergers at late epochs are either both
early-type (and therefore reasonably gas-poor) or that at most one
progenitor is a late-type galaxy. It is worth noting that recent
theoretical studies have shown that even major mergers involving
gas-rich progenitors are not particularly efficient at driving
significant star formation
\citep[e.g.][]{Kapferer2005,DiMatteo2007,DiMatteo2008,Cox2008,Robaina2009}.
However, the results of \citet{Kaviraj2009} indicate that the
bluest ETGs in the UV CMR are compatible with merger mass ratios
between 1:4 and 1:6. While this does suggest that major mergers
are unlikely to dominate the blue ETG population, the reader is
cautioned that a major-merger origin for some of the blue ETGs
cannot be completely ruled out.}

Since morphological disturbances {\color{black}do} typically
accompany UV blue colours, and there are too many disturbed ETGs
than can be accounted for by major mergers alone, our results
point towards minor mergers contributing the
{\color{black}\emph{bulk}} of the RSF in the ETG population at
late epochs, with at least 60\% (and perhaps $>90$\% if the
\citet{Lin2004} value is more representative) of the disturbed
ETGs having experienced a minor merger i.e. the accretion of a
small gas-rich satellite onto a massive spheroid.

{\color{black}It is worth noting that the scatter in UV/optical
colours of the ETG population persists over the last 8 billion
years \citep{Kaviraj2008b}. Since each RSF event is expected to
remain `visible' due to blue UV colours for $\sim1$ Gyr (after
which the UV decays rapidly), it is reasonable to suggest that
ETGs, on average, may experience several such (minor merger)
events over their lifetimes. The contribution from major mergers
is almost certainly larger at higher redshifts ($z>2$) and
plausibly declines at later epochs, although consensus on the
actual rate of that decline requires more investigation \citep[see
e.g.][]{Bundy2009}}.

{\color{black}As we briefly noted in the introduction} several
works in the recent literature, that have explored various aspects
of massive galaxy or specifically ETG evolution at late epochs,
support our findings and highlight the potentially significant
impact of minor merging on the evolution of the ETG population. In
a direct parallel to our conclusions, \citet{Naab2009} have used
hydrodynamical simulations to demonstrate that, even though major
mergers do occur, the evolution in the size and density profiles
of luminous early-type galaxies is likely to be driven by repeated
minor merging at late epochs ($z<2$). A similar result has been
determined from an observational viewpoint by
\citet{Bezanson2009}, albeit for a mass range that is higher than
the population studied here ($M>10^{11}$M$_{\odot}$).

Several efforts aimed at correlating the rate of major mergers
with the evolution of massive galaxies indicate that the observed
major merger rate at late epochs appears insufficient to satisfy
the evolution in the number densities of massive galaxies since
$z\sim1.2$ \citep{Bundy2007,Bundy2009} and that the contribution
of major merger activity to the cosmic star formation rate is
small \citep[e.g.][]{Robaina2009,Jogee2009}. While environmental
processes may induce morphological transformations in regions of
high density \citep[see e.g.][and references
therein]{Kawata2008,Rasmussen2008}, the paucity of major mergers
may indicate that the buildup of the field ETG population (as
studied here) is significantly influenced by minor merging, since
these events are capable of producing remnants that resemble
present-day early-type galaxies \citep{Naab2009,Bournaud2007}.

The kinematic properties of early-types in our local neighbourhood
provide further clues on the impact of minor mergers. Results from
the SAURON project \citep{sauron2}, which has used integral field
spectroscopy to perform a systematic study of the stellar and gas
kinematics in a representative sample of 48 early-types in the
local Universe, indicates that 75\% of early-types are `fast
rotators'\footnote{This value is probably closer to 90\%, based on
results from the ATLAS$^{\textnormal{3D}}$ project, which is the
successor to SAURON and based on a complete sample of $\sim250$
ETGs within the local 40 Mpc volume (Emsellem et al. 2010, in
prep.; Roger Davies and Martin Bureau, priv. comm.). The survey
homepage can be found here:
http://www-astro.physics.ox.ac.uk/atlas3d/} i.e. their stellar
velocity fields show a clear large-scale rotation pattern. Since
stellar disks are easily perturbed in major mergers, even in the
presence of moderate amounts of gas \citep{Naab2006}, the
properties of the fast rotators indicate that the overwhelming
majority of the SAURON ETGs have experienced one or more minor
mergers up to $z=0$ \citep{Emsellem2007}.

While this study provides an explicit link between star formation
activity and minor mergers, taken together with the aforementioned
parts of the literature, it is reasonable to conclude that the
evolution in both structural properties and star formation
activity and consequently the overall evolution of the ETG
population at late epochs may be significantly influenced by minor
mergers.

%.........................................................................................................

\subsection{{\color{black}Faint morphological disturbances in the relaxed ETGs?}} {\color{black}We should note that the detection of
morphological disturbances depends on the surface brightness limit
of the images. The advantages of deeper imaging have been revealed
in a recent study by \citet[][VD2005 hereafter]{VD2005}, who used
very deep optical imaging (to a limiting surface brightness $\mu
\sim 28$ mags arcsec$^{-2}$) to study local galaxies on the
optical red sequence. VD2005 found widespread signatures of recent
merging in a large fraction (70\%) of their ETG sample, which are
invisible in shallower imaging from e.g. the SDSS survey
\citep[see also][]{Tal2009}. It is likely, therefore, that many of
the galaxies that are classified as relaxed ETGs (regardless of
their UV colour) harbour tidal disturbances that are fainter than
the surface brightness limit of the COSMOS images used in this
study.

For example, as we noted in Section 2.2, the visual classification
identifies some relaxed ETGs in the UV blue cloud (see objects
1618, 3156 and 3796 in Figure \ref{fig:etgrelaxed_examples}).
Object 3796 might have a slightly elongated core compared to the
other relaxed ETGs, suggesting that it may be in the very last
stages of a recent merger. Similarly, there is a hint that object
1618 may exhibit some low surface brightness disturbances in the
north western part of the galaxy which may be more readily visible
in deeper imaging. While the putative features in these galaxies
are too faint for us to confidently classify them as disturbed
ETGs, it is reasonable to suggest that many of the relaxed ETGs
(regardless of whether they have red or blue colours) might
harbour unseen morphological disturbances that are too faint to be
detected in the COSMOS images.}

%.........................................................................................................

\subsection{Alternative drivers of star formation in ETGs?}
While our results present a strong case for the RSF in the ETG
population to be driven by merging activity, it is instructive to
study other potential sources of gas that might contribute to the
observed star formation.

A potential source of cold gas comes from internally recycled
material, injected into the interstellar medium (ISM) through
stellar winds and supernovae. Up to 30\% of the IMF could be
expected to return over the lifetime of a stellar population
\citep[e.g.][]{Ferreras2000b} and some or all of this gas could be
expected to fuel star formation. However, \citet[][see their
Section 6]{Kaviraj2007b} have suggested that the recycled gas mass
at late epochs is almost an order of magnitude too low to produce
early-type galaxies in the blue cloud. This is supported by
detailed spatial studies of (ionised) gas in ETGs using integral
field spectroscopy in the SAURON project \citep[e.g.][]{sauron5},
which indicate that the kinematics of the gas is typically
decoupled from the stellar field, indicating, at least in part, an
external origin. {\color{black}Crucially, RSF caused by internally
recycled gas would not alter the morphology of the galaxy. Hence,
in the context of this study, if the RSF was indeed driven by this
channel, one would expect many relaxed ETGs in the UV blue cloud,
something that is clearly not observed. Therefore, it is
reasonable to suggest that internally recycled gas is not the
primary driver of RSF in the ETG population at intermediate
redshifts.}

A second possible route to driving star formation, without
recourse to mergers, is through condensation of gas from hot gas
reservoirs (the galactic equivalent of a `cooling flow'). This
mechanism will only operate in the most luminous ETGs
(specifically the central galaxies in the central of large
potential wells such as clusters), since their moderate and low
luminosity counterparts do not have hot gas envelopes
\citep[e.g.][]{Fabbiano1989,Mathews2003}. However, the
\emph{total} mass of hot gas in massive ETGs typically forms a
small fraction (a few percent or less) of the stellar mass of
these systems
\citep[e.g.][]{Bregman1992,OSullivan2003,Mathews2003}. Recalling
that the RSF mass fractions are themselves around $\sim10$\% or
less and form within a few tens of Myrs at most, it is difficult
to envisage a substantial fraction of the hot gas reservoir
cooling over such timescales to produce the observed RSF.
{\color{black}Furthermore, as in the argument against internally
recycled gas, one would expect condensation-driven RSF to leave
the morphology of the galaxy unchanged, so that we would see a
reasonably large number of relaxed ETGs in the UV blue cloud,
which is not observed.}

{\color{black}Finally, it should also be noted that some
contribution from cold flows
\citep[e.g.][]{Keres2005,Birnboim2003,Birnboim2007} cannot be
completely ruled out. Numerical simulations indicate that the
contribution of cold flows is largest in the high-redshift
Universe. Accretion of gas through filaments onto central galaxies
is seen to become inefficient at $z\sim 4$, while the process
continues to affect satellite galaxies in haloes until $z \sim 2$,
with mergers becoming the dominant process driving star formation
thereafter \citep{Khalatyan2008}. As with the mechanisms discussed
above, star formation driven by filamentary cold flows would not
induce strong morphological disturbances in the main body of the
galaxy. However, while the vast majority of disturbed ETGs in our
sample show clearly disturbed morphologies, some systems, e.g.
3074 in Figure 3, exhibit a mild morphological disturbance in the
main body of the galaxy but may show some evidence for asymmetric
gas accretion. Given the redshift of our galaxy sample ($z \sim
0.6$) and the fact that the disturbed ETGs typically carry
significant morphological disturbances, cold flows are unlikely to
be the main driver of RSF in ETGs at these epochs. Nevertheless,
their contribution to some of the star formation cannot be
completely ruled out.}

%.........................................................................................................

\section{Summary and conclusions}
We have exploited rest-frame UV and optical photometry of
intermediate-redshift ($0.5<z<0.7$) early-type galaxies (ETGs) in
the COSMOS survey to demonstrate that the low-level star formation
activity in these systems is likely to be dominated by minor
mergers. $z\sim0.5$ is the closest redshift at which the COSMOS
$u$-band filter traces the rest-frame near-UV (NUV) spectrum
($\sim2500$\AA). Selecting galaxies as near as possible also
maximises the accuracy of eyeball morphological classifications
and minimises morphological $k$-corrections. Furthermore, placing
the study at redshifts greater than $z\sim0.5$ ensures that the UV
flux in the early-type population comes exclusively from young
stars. Potential contributions to the UV from old horizontal
branch stars and their progeny do not exist at these redshifts
because the Universe is not old enough for the HB to have
developed.

Visual inspection of the COSMOS HST images has been used to split
the ETG population into galaxies which show morphological
signatures of recent merging and those that are relaxed in the
COSMOS HST images. The disturbed ETGs form $\sim32$\% of the ETG
population. The relaxed ETGs are almost completely contained
within the UV red sequence ($u-i>3$). At luminosities greater than
$M_V\sim-22$, less than 5\% of relaxed ETGs occupy the UV blue
cloud, while at fainter luminosities the fraction residing in the
UV blue cloud rises marginally to $\sim$10\%. In contrast, the
morphologically disturbed ETGs favour the UV blue cloud at all
luminosities. At luminosities greater than $M_V\sim-22$, more than
50\% of the disturbed ETGs reside in the UV blue cloud ($u-i<3$),
while at lower luminosities the fraction is substantially higher
($\sim$85\%). {\color{black}This abrupt change in the blue
fraction at lower luminosities may be partly affected by
incompleteness on the UV red sequence.} At a given luminosity, the
disturbed ETGs are at least 1.5 mags bluer in their rest-frame UV
colours than their relaxed counterparts. Most importantly, at all
luminosities, the scatter to blue colours in the early-type
population is dominated by the disturbed ETG population. The
recent star formation (RSF) in the disturbed ETGs typically has
ages between $\sim0.03$ Myrs and $\sim0.3$ Myrs and contributes
less than 10\% of the mass fraction of the remnants.

{\color{black}The general lack of relaxed ETGs in the UV blue
cloud argues against the RSF being driven by either internal mass
loss or condensation from hot gas reservoirs, since these
processes are not expected to perturb the morphology of the
galaxies.} The preponderance of disturbed ETGs in the UV blue
cloud indicates that merging plays a dominant role in driving the
star formation activity in these galaxies.
{\color{black}Empirically and theoretically derived major merger
rates} suggest that the frequency of major merging in galaxies
with masses similar to the ones considered here is too low to
fully account for the significant fraction of disturbed ETGs.
While major mergers play a lesser role, minor mergers between
massive spheroids and small gas-rich satellites are likely to be
responsible for at least 60\% (and possibly $>90$\%) of the
disturbed ETG population at intermediate redshift.
{\color{black}As we noted in Section 5 above, the large scatter in
the UV/optical colours and associated star formation in the ETG
population persists over the redshift range $0<z<1$, which
corresponds to $\sim$ 8 Gyrs in look-back time. Since the UV
signatures of any given RSF event is expected to decay after 2
Gyrs or less, it is reasonable to suggest that ETGs, on average,
may experience several minor merger events over their lifetimes.
Since several studies have suggested that the evolution in the
structural properties of ETGs are also consistent with minor
merging, we have argued that our results, taken together with the
recent literature, suggests that the overall evolution of the ETG
population over the last 60\% of the lifetime of the Universe may
be significantly influenced (or perhaps even driven) by minor
mergers.}

Despite recent interest \citep[e.g.][and this
paper]{Jogee2009,Lopez2010}, minor mergers remain a largely
unexplored topic in the modern galaxy evolution effort. Pair
studies, on which much of our understanding of merging is based,
require a spectroscopic detection of both progenitors in a galaxy
pair, since the relative velocity is part of the pair criterion.
The population of pairs that can be extracted from a particular
survey therefore depends critically on the spectroscopic limit of
the survey in question. While major mergers are easily identified,
minor mergers are harder to detect because the smaller progenitor
is often fainter than the spectroscopic limit of the survey. The
accumulating evidence for the significance of minor mergers calls
for deep spectroscopic surveys in the future that would allow
statistical studies of minor merging to be conducted, so as to
quantify the impact of this potentially important process on the
evolution of the galaxy population, especially in the redshift
range $0<z<1$.

%.........................................................................................................

\section*{Acknowledgements}
The anonymous referee is gratefully acknowledged for a thorough
report which significantly improved the original manuscript. SK
acknowledges an Imperial College Junior Research Fellowship, a
Research Fellowship from the Royal Commission for the Exhibition
of 1851, a Senior Research Fellowship from Worcester College,
Oxford and support from the BIPAC institute at Oxford. Some of
this work was supported by a Leverhulme Early-Career Fellowship
(SK). {\color{black}RSE acknowledges support from the Royal
Society}. Roger Davies, Kevin Bundy, Martin Bureau, Chris
Conselice, Jerry Sellwood and Sukyoung Yi are thanked for
interesting discussions and useful comments. Some of this research
utilised the undergraduate computing facilities in the Department
of Physics at the University of Oxford.

%.........................................................................................................

\nocite{Han2003}

%.........................................................................................................

\bibliographystyle{mn2e}
\bibliography{references}

%.........................................................................................................

\end{document}